\documentclass[useAMS,usenatbib]{mn2e}
\usepackage[dvips]{lscape,graphicx}
\usepackage{url}
\usepackage{amssymb,amsmath}
\usepackage{subfigure}

\def\gtabouteq{\,\hbox{\raise 0.5 ex \hbox{$>$}\kern-.77em 
                    \lower 0.5 ex \hbox{$\sim$}$\,$}}       
\def\ltabouteq{\,\hbox{\raise 0.5 ex \hbox{$<$}\kern-.77em 
                     \lower 0.5 ex \hbox{$\sim$}$\,$}}

\title[CO(J=3-2) Observations of NGC~4631]{The JCMT Nearby Galaxies Legacy Survey V: The CO(J=3-2)
Distribution 
and Molecular Outflow in NGC~4631}
\author[J. A. Irwin et al.]{Judith A. Irwin$^{1}$,
C. D. Wilson$^{2}$,
T. Wiegert$^{3}$
G. J. Bendo$^{4}$,
B. E. Warren$^{2,5}$,
\newauthor 
Q. D. Wang$^{6}$,
F. P. Israel$^{7}$, 
S. Serjeant$^{8}$, J. H. Knapen$^{9,10}$, E. Brinks$^{11}$, 
\newauthor
R. P. J. Tilanus$^{12}$, P. van der Werf$^{13}$ and S. M{\"u}hle$^{14}$\\
$^{1}$Dept. of Physics, Eng. Physics, \&
Astronomy, Queen's University, Kingston, Ontario, K7L 3N6, Canada,
irwin@astro.queensu.ca\\
$^{2}$Dept. of Physics \& Astronomy, McMaster University, Hamilton,
Ontario, L8S 4M1, Canada, wilson@physics.mcmaster.ca\\
$^{3}$Dept. of Physics \& Astronomy, University of Manitoba, Winnipeg,
Manitoba, R3T 2N2, Canada, wiegert@physics.umanitoba.ca\\
$^{4}$Astrophysics Group, Imperial College London, Blackett Laboratory,
Prince Consort Road, London SW7 2AZ, g.bendo@imperial.ac.uk\\
$^{5}$International Centre for Radio Astronomy Research, M468, 
University of Western Australia, Crawley, WA, 6009
Australia,\\ bradley.warren@icrar.org\\
$^6$Department of Astronomy, University of Massachusetts, B-524, LGRT, Amherst, 
MA 01003, wqd@astro.umass.edu\\
$^{7}$Sterrewacht Leiden, Leiden University, PO Box 9513, 2300 RA Leiden,
The Netherlands, israel@strw.leidenuniv.nl\\
$^{8}$Dept. of Physics \& Astronomy, The Open University, Milton Keynes,
MK7 6AA, England, s.serjeant@open.ac.uk\\
$^9$Instituto de Astrof{\'i}sica de Canarias, E-38200 La Laguna, Tenerife, Spain,
jhk@iac.es\\
$^{10}$Departamento de Astrof{\'i}sica, Universidad de La Laguna, E-38205 La Laguna, Tenerife, Spain\\
$^{11}$Centre for Astrophysics Research, Science and Technology Research Institute, University of 
Hertfordshire, Hatfield AL10 9AB, UK,\\
 e.brinks@herts.ac.uk\\
$^{12}$Joint Astronomy Centre, 660 N. Aâohoku Pl., Hilo, Hawaii, 96720, USA, r.tilanus@jach.hawaii.edu\\
$^{13}$ Leiden University, Leiden Obsevatory, PO Box 9513, 2300 RA Leiden,
The Netherlands, pvdwerf@strw.leidenuniv.nl\\
$^{14}$Joint Institute for VLBI in Europe, Postbus 2, 7990 AA Dwingeloo, The Netherlands, muehle@jive.nl\\
}
\begin{document}

\date{Accepted year month day. Received year month day; in original form year month day}

\pagerange{\pageref{firstpage}--\pageref{lastpage}} \pubyear{2008}

\maketitle

\label{firstpage}


\begin{abstract}
We have made the first map of CO(J=3-2) emission covering the disk of the edge-on
galaxy, NGC~4631, which is known for its spectacular gaseous halo.
 The strongest emission, which we model with a Gaussian ring,
 occurs within a radius of
5 kpc.  Weaker disk emission
is detected out to radii of 12 kpc, the most extensive molecular component yet
seen in this galaxy. From comparisons with infrared data, we find that CO(J=3-2)
emission more closely follows the hot dust component, rather than the cold dust,
consistent with it being a good tracer of star formation. The first maps
of $R_{3-2/1-0}$, H$_2$ mass surface density 
and SFE have been made for the inner 2.4 kpc radius region.  Only 20\% of the SF
occurs in this region and excitation
conditions are typical of galaxy disks, rather than of central starbursts. The SFE
suggests long gas consumption timescales ($>$ $10^9$ yr).

The velocity field is dominated by a steeply rising rotation curve in the region of
the central molecular ring followed by a flatter curve in the disk.
A very steep gradient in the rotation curve is observed at the nucleus, providing the first
evidence for a 
central concentration of mass:
M$_{dyn}\,=\,5\,\times\,10^7$ M$_\odot$ within a radius of 282 pc.
The velocity field shows anomalous features indicating the presence of molecular outflows;  
one of them is
associated with a previously observed CO(J=1-0) expanding shell.  Consistent with these outflows
is the presence of a thick ($z$ up to $1.4$ kpc)
 CO(J=3-2) disk. 
  We suggest that the interaction between NGC~4631 and
its companion(s) has agitated the disk and also initiated star formation which was likely
higher in the past than it is now.   These may be necessary conditions 
for seeing prominent halos.

\end{abstract}

\section{Introduction}
\label{sec:intro}

\begin{figure*}
\includegraphics[scale=1]{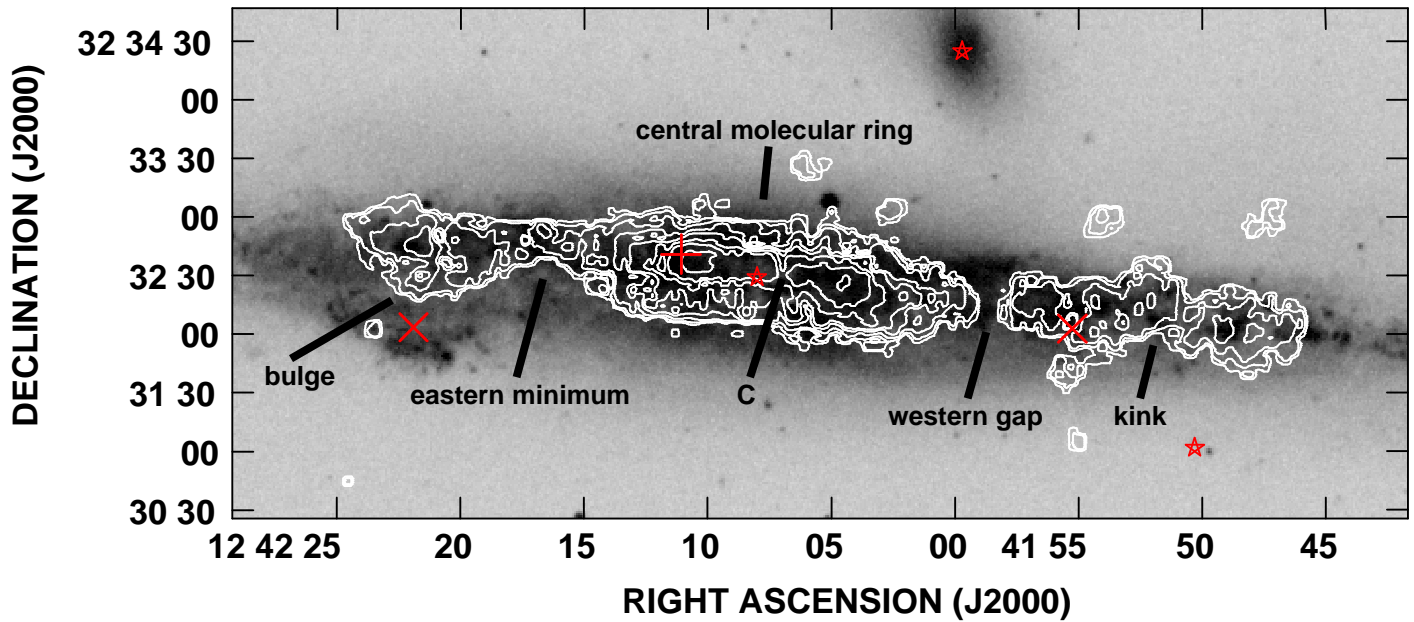}
\caption{CO(J=3-2) integrated intensity (zeroth moment) map of NGC~4631 superimposed on
the Second Digitized Sky Survey (DSS2) blue image.  The CO(J=3-2) field of
view shown here is 9.8$^\prime$ $\times$ 4.1$^\prime$ in
right ascension and declination, respectively. Contours are at
0.25, 0.75, 2.0, 5.0, 8.0, 15, 24, and 45 K km s$^{-1}$ and the peak
value is 57.7 K km s$^{-1}$. The moment-generating routine applied a
spatial gaussian 
smoothing function of 40$^{\prime\prime}$ FWHM and a cutoff level of
0.018 K before integrating over the unsmoothed cube from
437 km s$^{-1}$ to 821 km s$^{-1}$ (see text).
Red stars denote the centers of NGC~4631 (Table~\ref{tab:galaxy_params}),
the dwarf elliptical
companion, NGC~4627 
to the north-west, and the optical dwarf
galaxy candidate, NGC~4631 Dw A, to the south-west. 
The red plus (+) and two crosses (X) denote the
approximate centers of the CO(J=1-0) shell found by 
Rand (2000a) and the two HI supershells found by
Rand \& van der Hulst (1993, their Fig. 3), respectively. 
Features discussed in
Sect.~\ref{sec:results:distribution} are labelled.}
\label{fig:mom0_optical}
\end{figure*}

NGC~4631 (Fig.~\ref{fig:mom0_optical}, Table~\ref{tab:galaxy_params}) 
is an edge-on\footnote{We take `edge-on' to mean an inclination
greater than 85$^\circ$.} galaxy that is known for
a spectacular
 multi-phase halo\footnote{The term, `halo' is used to mean any extraplanar gas
or dust, where we conservatively take `extraplanar' to
imply $z\,\gtabouteq\,1$ kpc .}. 
This galaxy is one of the targets of the
James Clerk Maxwell Telescope (JCMT)
Nearby Galaxies Legacy Survey 
(NGLS)\footnote{http://www.jach.hawaii.edu/JCMT/surveys/}
\citep{wil09, war10} whose goals include  searching for molecular gas
and dust in nearby galaxies and comparing the global properties of such
systems.  In addition, 
the spatial and spectral coverage of the
325 - 375 GHz band presented by the new 
 Heterodyne Array Receiver Programme - B-band (HARP-B) detector
 together with the wide-band
Auto-Correlation Spectrometer Imaging System (ACSIS) has also made it possible
to study individual galaxies in the sample 
in some detail.  For NGC~4631, our
goals are to examine the CO(J=3-2)
properties and distribution in this unique galaxy and to relate, where
possible, the molecular emission to known outflow features.

The strength, prevalence, and multi-phase aspects of the halo of NGC~4631
make this galaxy unique and an important target for disk-halo studies.
The halo is observed in all ISM components, including
cosmic rays (CRs) and magnetic fields as indicated by
polarized radio continuum emission
\citep{gol94a,hum91,str04a}, HI \citep{ran94,ran96}
diffuse ionized gas \citep{mar01,ran92,ott03,hoo99},
dust \citep{nei99,mar01},
molecular gas \citep{ran00a}, and
hot, X-ray emitting gas 
\citep{vog96,tul06a,tul06b, wan01,wan95,str04a,yam09}. 
Expanding shells have also been observed in HI \citep{ran93} and
CO(J=1-0) \citep{ran00a}.
The halo extends over the entire star forming disk,
reaching a variety of 
vertical heights, $z$, depending
on the component considered, from\footnote{We adjust all values to a common adopted distance of
9 Mpc (Table 1), for comparative purposes.}
 900 pc in molecular gas
\citep{ran00a} to 10 kpc in the radio continuum 
\citep{gol94a}.  

NGC~4631 is interacting with two
companions, a dwarf elliptical, NGC~4627, 
$2.6^\prime$ (6.8 kpc, $\Delta V_{sys}\,=\,64$ 
km s$^{-1}$)\footnote{$\Delta V_{sys}\,\equiv\,V_{sys}(NGC4631) - 
V_{sys}(companion)$.}
 to the north-west (Fig.~\ref{fig:mom0_optical})
and
NGC~4656, another large edge-on galaxy about $32^\prime$ 
(84 kpc, $\Delta V_{sys}\,=\,-40$ km s$^{-1}$) to the south-east,
the result being 4 long intergalactic HI streamers 
\citep{wel69,wel78, ran94}
stretching to $\approx\,$42 kpc.  The bases of these tidal streamers overlap
with the halo of NGC~4631.
In addition, three more faint companions have been detected in HI 
\citep{ran96,ran94} as well as 
a faint optical dwarf
galaxy candidate, NGC~4631 Dw A,
 $2.5$ kpc below the plane of NGC~4631 
\citep{set05a} for which no redshift data are yet available. 
Of these companions, two (NGC~4627 and NGC~4631 Dw A)
fall within the field shown in
Fig.~\ref{fig:mom0_optical} and  
are marked with stars.
Presumably, the star formation (SF) activity (and hence the halo)
in NGC~4631  has been triggered and/or enhanced by
interactions. 
Interactions have also likely produced the observed
 thick stellar disk, i.e. the optical emission
shows $sech^2\left(z/z_0\right)$
scale heights, $z_0$, up to 1.4 kpc (depending on the stellar population
considered) with 
 detections to many scale heights in $z$
\citep{set05b}.


Studies of the mid-IR emission and dust properties in NGC~4631 can
be found in \cite{dra07}, \cite{dum04}, \cite{smi07},
\cite{ste05}, \cite{ben03, ben06}, and \cite{dal05, dal07}.
Previous CO observations in lower $J$ transitions have been carried out by
\cite{pag01}, 
\cite{gol94b}, 
 \cite{ran00a},
\cite{tay03},
 and \cite{isr09}.  
\cite{isr09} also obtained a CO(J=3-2) measurement in a single beam at the center
of the galaxy.
Limited
previous CO(J=3-2) mapping has been carried out by \cite{dum01} who detected
emission only in the
inner, 2.6$^\prime$ diameter region 
with a spatial resolution of 
$\approx\,23^{\prime\prime}$.  The data presented here are of both higher
resolution and sensitivity and, as will be shown, reveal the distribution of
CO(J=3-2) both in the central regions as well as throughout the disk of
NGC~4631.  

Sect.~\ref{sec:obs_datared} outlines the observations and data reductions.  In
Sect.~\ref{sec:results} we describe the CO(J=3-2) distribution, 
and will be considering the distribution
in the disk and a comparison with other wavebands, the CO(J=3-2) excitation, molecular mass and
star formation, the velocity distribution and anomalous velocity and high latitude emission.
Sect.~\ref{sec:discussion} presents the discussion and Sect.~\ref{sec:conclusion} the conclusions.

\begin{table}
 \centering
 \begin{minipage}{140mm}
  \caption{Basic Parameters of NGC~4631}
  \label{tab:galaxy_params}
  \begin{tabular}{ll}
\hline
parameter & value \\
\hline
Hubble type\footnote{\cite{but07}.} & SB(s)d sp or Sc sp \\
RA (J2000) (h m s)$^b$ & 12 42 08.01  \\
DEC (J2000) ($^\circ$ $^\prime$ $^{\prime\prime}$)\footnote{IR center at $2\,\mu$m from
the Nasa Extragalactic Database (NED).}
&  32 32 29.4  \\
$V_{sys}$ (km s$^{-1}$)\footnote{Systemic HI velocity  
(heliocentric, optical definition) from NED.} & 606 \\
$D$ (Mpc)\footnote{Distance \citep[e.g.][]{ken03}.} 
& 9.0  \\
a $\times$ b ($^\prime$
$\times$ $^\prime$)\footnote{Optical major $\times$
minor axis, measured to
 the 25 blue mag\\ per square arcsec brightness level \citep{dev91}.} & 15.5 
$\times$ 2.7\\
$PA$\footnote{Position angle \citep{dev91}.} & 86$^\circ$\\
$i$\footnote{Adopted inclination \citep[e.g.][]{isr09}.} &  86$^\circ$\\
$L_{FIR}$ (erg s$^{-1}$)\footnote{Far-IR luminosity \citep{tul06a}, 
adjusted to\\
$D\,=\,9.0$ Mpc.} 
& $9.5\,\times\,10^{43}$\\
$L_{FIR}/D_{25}^2$ 
($10^{40}$ erg s$^{-1}$ kpc$^{-2}$)\footnote{From
\cite{tul06a}, where $D_{25}$ is the galaxy's\\
diameter measured at the 25 blue mag per square arcsec level.}
& 7.76 \\
SFR$_{FIR}$ (M$_\odot$ yr$^{-1}$)\footnote{Far-IR SFR \citep{tul06a}. See also Sect.~\ref{sec:results:star_formation}.} 
& 4.3\\
SFR$_{H\alpha,corr}$ (M$_\odot$ yr$^{-1}$)\footnote{H$\alpha$ SFR corrected for
extinction according to the formula of\\ \cite{cal07}, 
but using the H$\alpha$ correction of\\ \cite{zhu08}; the
H$\alpha$ luminosity of \cite{hoo99}\\ and $\lambda\,24~\mu m$ flux of
 \cite{dal07} have been used. 
} 
& 2.4\\
$M_{HI}$ (M$_\odot$)\footnote{HI mass \citep{ran94}.} &  
$1.0\,\times\,10^{10}$\\
$M_{dust}$ (M$_\odot$)\footnote{Dust mass
\citep{ben06}.} 
& $9.7\,\times\,10^7$\\
\hline
\end{tabular}
\end{minipage}
\end{table}

\section{Observations and Data Reduction}
\label{sec:obs_datared}

Data were obtained of the $^{12}$CO(J=3-2) (rest frequency, $\nu_0$ = 345.7959899 GHz) spectral
line at the JCMT
 using the HARP-B front-end and the ACSIS back-end (Smith et al. 2003).
The HARP-B array  
contains 16 receivers in a 4 $\times$ 4 pattern separated by 30$^{\prime\prime}$
(about two beam-widths). In order to fully sample the field,
the observing was set up as a sequence of raster scans 
in a ``basket weave'' pattern so that 
scanning was carried out in both the major and minor axis directions.  The final
sampling spacing was 1/2 of the beam size.
Complete details of the observing set-up can be found in \cite{war10} and
Table~\ref{tab:observing}.
In total, 14 scans were obtained over two nights under good conditions 
with the
$\nu\,=\,225$ GHz optical depth,
$\tau_{225\,GHz}$, ranging from 0.051 to 0.070 on January 5 and from
0.041 to 0.068 on January 6.  The calibration sources were IRC+10216 and Mars.
Pointing offsets over the course of the observations were 2$^{\prime\prime}$ rms.

Data reduction was initially carried out using the Joint Astronomy Centre version
of the Starlink software\footnote{See http://starlink.jach.hawaii.edu
or \cite{cur08}.} using the {\it KAPPA},
{\it SMURF} and {\it CONVERT} packages for editing, cube making, and converting to 
flexible image transport system (FITS) format, respectively.
Visualization programs such as {\it GAIA} and {\it SPLAT} allowed us to inspect the data. 
Two of the 16 receptors were not functioning properly and had to be removed from all data.
 The editing was iterative, beginning by inspecting cubes from each scan
individually, removing
end channels, removing obvious interference spikes, binning to 20 km s$^{-1}$, removing
a linear baseline, and then collapsing the cube to inspect the
total intensity (zeroth moment) map.  This usually
revealed pixels that had obviously poor baselines for any given scan.  Poor data points were removed from the
unbinned, unbaselined data, and the process repeated, as required.  All scans of the edited, but otherwise
original resolution 
data were then combined into a single cube, using a `SincSinc' kernel\footnote{The
{\it MAKECUBE} routine in {\it SMURF}
was used.} with a cell size of 3.638 arcsec (1/4 of the beam)
 and then saved in FITS format.  


The fits cube was then read
into the Astronomical Image Processing System (AIPS) package for the remainder of the processing
and analysis. 
 The data were box-car binned to a velocity resolution of 
10.4 km s$^{-1}$ and 
a linear baseline was removed, fitted pixel by pixel.  Some further minor editing was then
carried out in AIPS.  In addition, residual baseline curvature was still
evident in some sections of the cube
 and these were then flattened using a 3rd order polynomial. 

\begin{figure*}
\includegraphics[scale=1]{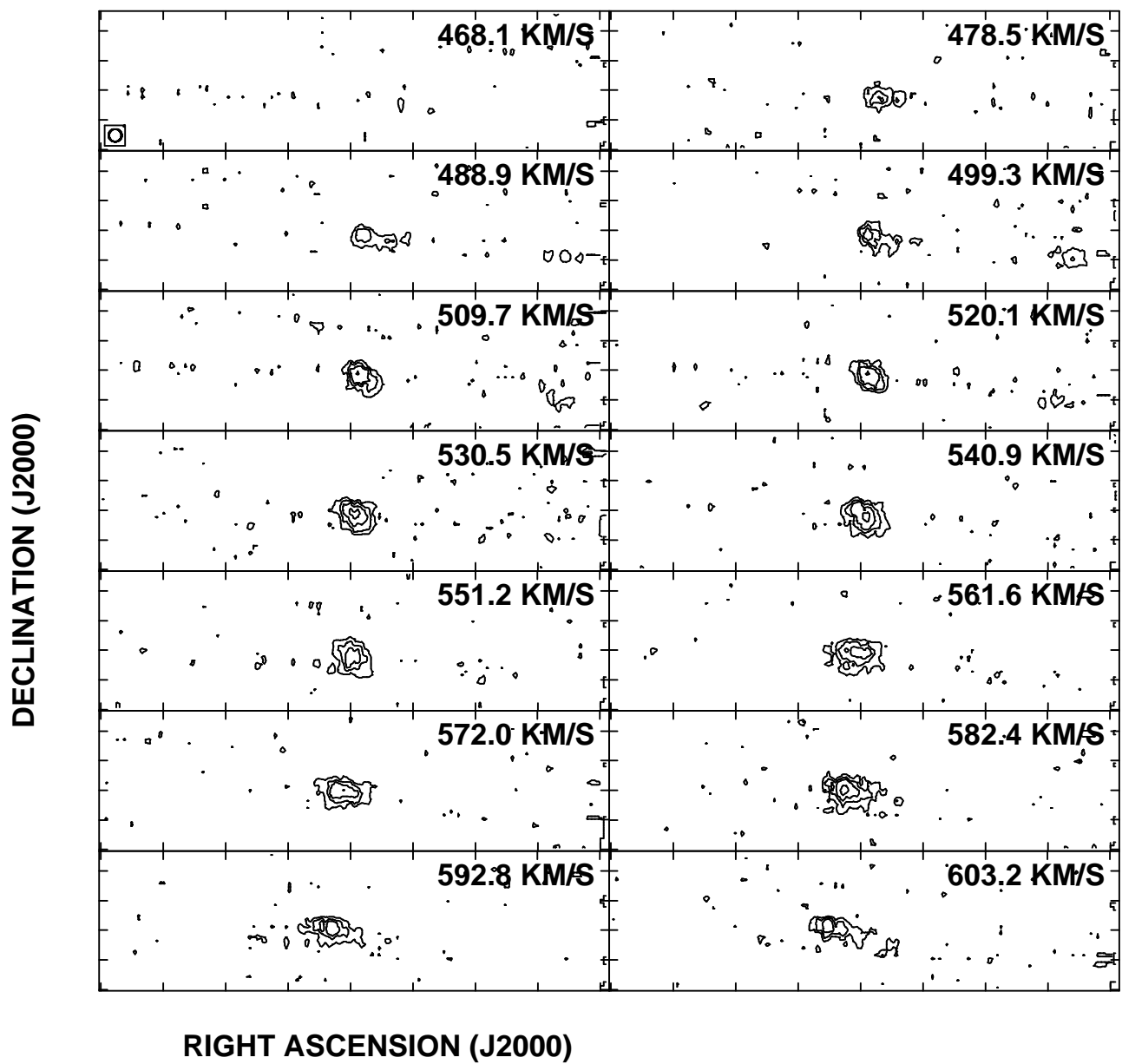}
\caption{{\bf (a)} CO(J=3-2) channel maps.  Contours are at 0.10 
($3\,\sigma$), 0.20, 0.30,
0.50, and 0.75 K.  The declination and right ascension scales are shown in
Fig.~\ref{fig:chanmapsb}b.  The velocity of each channel is given at upper right and
the beam is shown at the lower left of the first frame.
}
\label{fig:chanmapsa}
\end{figure*}

The final data were subsequently corrected for the main beam efficiency, 
$\eta_{MB}=0.60$ (estimated uncertainty between
 10 and 15\%) in order to
convert into units of main beam brightness temperature, T$_{MB}$.
Final channel maps for those channels that display emission are shown
in Fig.~\ref{fig:chanmapsa}a and b.
The resulting measured
rms noise (Table~\ref{tab:observing}) per channel, 
met our goal for the NGLS\footnote{The target rms per 20 km s$^{-1}$ channel
was 0.030 K (T$_{MB}$) which matches our measured value with binning to the wider channel.  Note, however, that
the noise increases towards the map edges.}.
An examination of each individual HARP beam from independent observations
of Mars shows that sidelobes of order 3\% of the peak occur at distances
of 24$^{\prime\prime}$ from the beam center which is well within our estimated
uncertainties (see below). Sidelobes at   
 larger distances are estimated to be less
than this (P. Friberg, private communication) and therefore negligible.

A total intensity (zeroth moment)
 map was then made, 
which
 involved smoothing the data cube spatially, 
imposing a flux cutoff based on the smoothed cube, 
and then integrating in velocity over the original
cube, using only those pixels that, in the smoothed cube, were above the flux
cutoff. 
The result is shown in
Fig.~\ref{fig:mom0_optical}, with cutoff and smoothing details given in the caption. 
The field of view has also been reduced slightly (to 80\% of the original)
 in order to trim noisy edge points.

The noise, taken to represent the uncertainty due to random errors
 in our total intensity map, is 
$\approx\,$1.6 K km s$^{-1}$, based on the noise per channel
and number of channels entering into the sum
at any typical position containing emission.  In addition, we
can compare our map to one obtained independently from the
same data but using
different software, editing, pixel size, channel width, and
baseline smoothing \citep[see][]{war10}.  The maxima of the two
maps differ
by 4\%, and a histogram of the difference map is well fit by
a Gaussian with a peak at 0.20 K km s$^{-1}$ and a standard deviation
of 1.0 K km s$^{-1}$.  These differences, the largest of
which are likely due to differences in baseline flattening, are less than our estimated
uncertainty above.
Aside from these random errors, an absolute calibration
error of 10 to 15\% is present due to uncertainties in the value of $\eta_{MB}$;
this uncertainty affects all pixels and does not change the appearance of
the map.  We have also compared the integrated intensity from the center of
Fig.~\ref{fig:mom0_optical} to
the single value given by \cite{isr09} who also used the
JCMT but
with a different receiver.  Our mean value of 
24 $\pm$ 3 K km s$^{-1}$
in a 14.5$^{\prime\prime}$
beam agrees with his result of 22 $\pm$ 3 K km s$^{-1}$ in a
 14$^{\prime\prime}$ beam at the same central position. 
%


The ancillary data used in this paper were taken from 
the Spitzer Infrared Nearby Galaxies Survey (SINGS)
Ancillary Data 
Archive\footnote{http://irsa.ipac.caltech.edu/data/SPITZER/SINGS} unless
otherwise indicated.  The archive contains primarily Spitzer data but also
includes ancillary images such as the H$\alpha$ image used in this paper.
More information about the SINGS program can be found in \cite{ken03}.

\begin{table}
 \centering
 \begin{minipage}{140mm}
  \caption{Observing \& Map Parameters of NGC~4631}
  \label{tab:observing}
  \begin{tabular}{ll}
\hline
parameter & value\\
\hline
Observing Date &  Jan. 05 \& 06, 2008\\
Total bandwidth & 1 GHz \\
Original channel width & 0.488 MHz (0.43 km s$^{-1}$)\\
Velocity-binned channel width & 11.7 MHz (10.4 km s$^{-1}$) \\
Angular resolution\footnote{Average of HARP beams.} & 14.5$^{\prime\prime}$ \\
rms (T$_{MB}$) per channel\footnote{For
 10.4 km s$^{-1}$ channel width.} & 0.034 K\\
\hline
\end{tabular}
\end{minipage}
\end{table}

\setcounter{figure}{1}
\begin{figure*}
\includegraphics[scale=1]{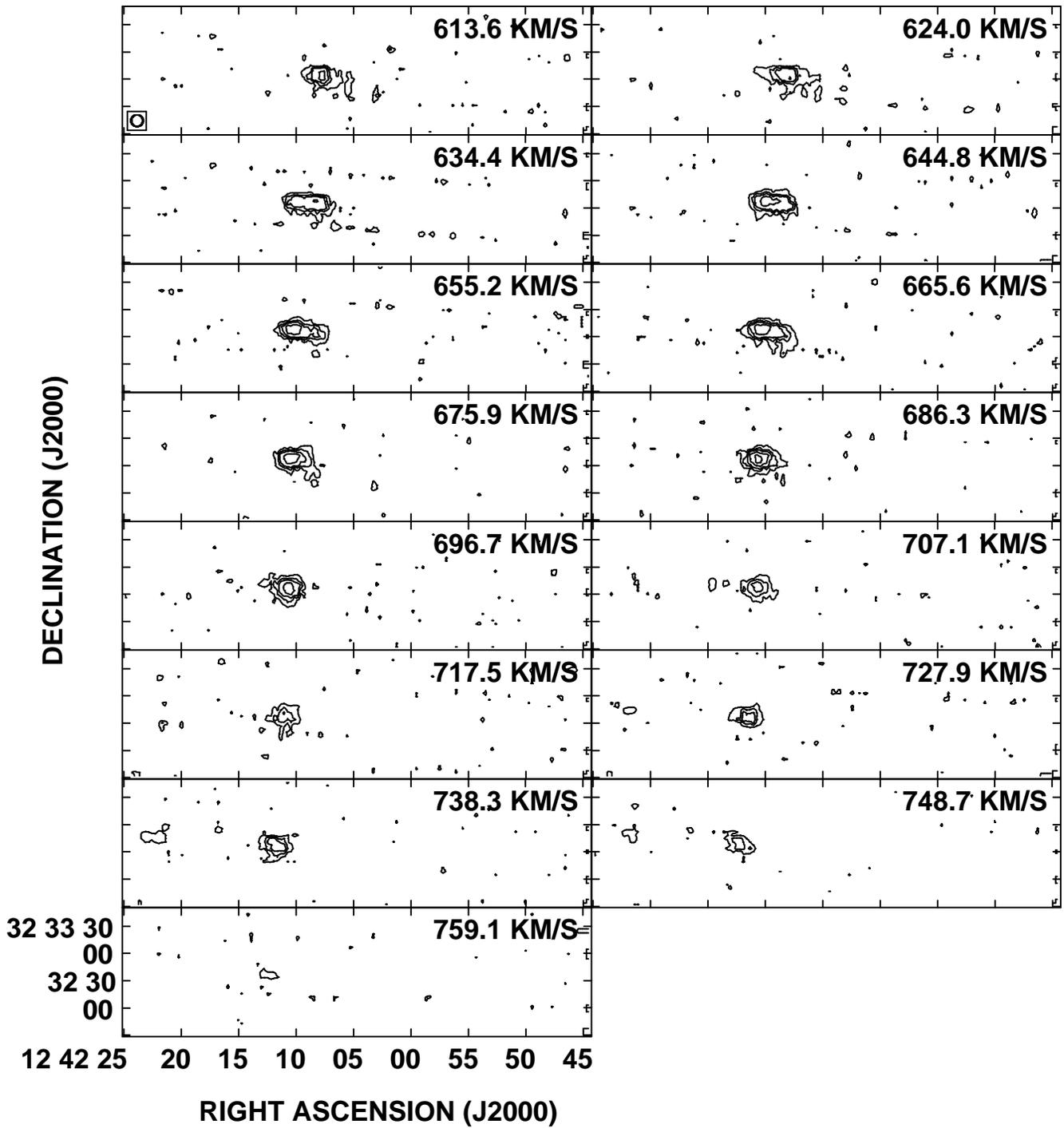}
\caption{{\bf (b)} CO(J=3-2) channel maps as in Fig.~\ref{fig:chanmapsa}a.}
\label{fig:chanmapsb}
\end{figure*}

\section{Results}
\label{sec:results}

\subsection{The CO(J=3-2) Distribution}
\label{sec:results:distribution}


As shown in the channel maps (Fig.~\ref{fig:chanmapsa}) the east side
of the galaxy is receding and the west side is approaching.  Both the
channel maps and the
total intensity map 
(Fig.~\ref{fig:mom0_optical}) show that the strongest emission
is concentrated in a region of diameter, $3.8^\prime$ 
(10.0 kpc) centered on the 
nucleus (labelled the `central molecular ring' in
Fig.~\ref{fig:mom0_optical} for reasons outlined in the next subsection).
At larger radii, there is weaker more extended disk emission. 
Several other features which we refer to below are also labelled
in Fig.~\ref{fig:mom0_optical}.
Fig.~\ref{fig:major_axis} shows a slice in emission along the
 major axis of
Fig.~\ref{fig:mom0_optical} at a position angle (83.5$^\circ$) 
chosen so that it passes
through the two broad maxima on either side of the nucleus (rather than 
the global optical major axis position angle of Table~\ref{tab:galaxy_params}).
Structures along the CO(J=3-2) major axis are well represented by this slice.
In the next subsections, we discuss features associated with the disk of NGC~4631;
discussion of the vertical distribution is deferred until
Sect.~\ref{subsec:results:faint}.

\subsubsection{The Strong Central Molecular Emission}
\label{sec:central_molecular_ring}

The strong central molecular emission is characterized by two peaks on either side
of the nucleus with a central minimum between them, and a slight
major axis curvature that is well-known in this galaxy.
This central emission extends between a minimum in the emission
on the east (the `eastern minimum') and a gap in the emission on the
west (the `western gap'), as labelled in Fig.~\ref{fig:mom0_optical}.
The eastern
peak is 15\% higher than the western
one, in agreement with the rudimentary 
CO(J=3-2) map of \cite{dum01}.  Since the 
CO(J=1-0) distribution \citep{gol94b, ran00a} 
also shows a central minimum, the observed central
 CO(J=3-2) minimum cannot be a result of lack of sufficient
excitation but must reflect a true minimum in the molecular
gas distribution between the two peaks.
A similar structure is also observed in dust emission 
\citep[e.g.][]{ben02, dum04, ben06, ste05};
we discuss comparisons with other wavebands in Sect.~\ref{sec:comparisons}.


Fig.~\ref{fig:major_axis} clearly shows that
this double-peaked central molecular emission dominates the 
 CO(J=3-2) distribution. 
The red curve 
represents
 an edge-on Gaussian ring with peak amplitudes that
are slightly asymmetric
 and whose fitted parameters 
are given in 
Table~\ref{tab:ring_model}.  
The emission was modelled following the
method of \cite{irw94} and \cite{irw96} which reproduces the 
line profiles of the cube. We
imposed a cutoff radius at $100^{\prime\prime}$ in order to model the strong
central emission only.
  The displayed model profile was then
obtained in the same way as the
data slice\footnote{The model is symmetric but the amplitude was arbitrarily fitted
and the slight asymmetry was reproduced with an insignificant adjustment of the position angle. The 
results of Table~\ref{tab:ring_model} are not affected by this.}. 
This Gaussian ring describes the central emission very well except for
 excess emission in the wings at larger radii and also
 some departures near the nucleus.  The inset shows the residuals.

The geometry of the central molecular gas distribution
may be more complex than a simple ring.  For example, 
bright spiral arms, rich in molecular
gas of the kind seen in more face-on galaxies such as M~51 \citep{bru08}, may
mimic a ring or pseudo-ring when observed edge-on.  
There has also been some suggestion that NGC~4631 is barred.
For example, 
there has been uncertainty in the optical classification (see Table~\ref{tab:galaxy_params})
and \cite{roy91} have suggested that the entire region that we have modelled as a ring
could be a large bar.  A molecular bar could also mimic the distribution shown in 
Fig.~\ref{fig:major_axis}, provided
 its density distribution is peaked at the bar ends.  While
this is possible, it has been shown \cite[see][and references therein]{kun07} that
barred galaxies are much more likely to show strong central peaks or 
concentrations of CO, rather than
the central minimum that we observe in NGC~4631.  
\cite{ran00a} also finds no evidence for a bar,
and a new high quality optical 
mosaic in {\em gri} colours from the Sloan Digital Sky Survey
 shows no clear bar in this 
galaxy either\footnote{See http://cosmo.nyu.edu/hogg/rc3/ courtesy of
David W. Hogg, Michael R. Blanton, and the Sloan Digital Sky Survey Collaboration.}.


In summary, other specific geometries could be invoked to explain the strong central
molecular emission, but since the Gaussian ring can do so with very few free parameters,
we will use the term `central molecular ring' to describe this region.
Its velocity distribution will be discussed in
Sect.~\ref{sec:velocity_ring} but it is worth noting here that
 the central molecular emission is
 kinematically distinct from the outer emission. 

\begin{table}
 \centering
 \begin{minipage}{140mm}
  \caption{Parameters of Modelled Central Molecular Ring}
  \label{tab:ring_model}
  \begin{tabular}{ll}
\hline
parameter & value\\
\hline
RA (J2000) (h m s)\footnote{Ring center position. Uncertainties indicate the variation\\
 that produces an estimated increase of 1$\sigma$ to the residuals.} & 12 42 7.7 $\pm$ 0.4 \\
DEC (J2000) ($^\circ$ $^\prime$ $^{\prime\prime}$)$^a$ & 32 32 30 $\pm$ 5 \\
$i$ (deg)\footnote{Best fit inclination.}& 89 $\pm$ 4  \\
$R_0$ ($^{\prime\prime}$, kpc)\footnote{Galactocentric radius of ring peak.} & 42 $\pm$ 3, 
1.8 $\pm$ 0.1 \\ 
$D_o$ ($^{\prime\prime}$, kpc)\footnote{Outer Gaussian scale length, i.e. 
$n(r)=n_0\,exp(-r^2/(2 D_o^2))$\\ where $n(r)$ is an in-plane density, $n_0$ is the density
at $R_0$, and $r$ \\ is a radial distance measured outwards from $R_0$.\label{footnote:inner} } 
& 6.4 $\pm$ 0.7, 0.28 $\pm$ 0.03  \\
$D_i$ ($^{\prime\prime}$, kpc)\footnote{Inner Gaussian scale length, as in 
Footnote~\ref{footnote:inner} with $D_i$\\ replacing $D_o$ and $r$ measured radially inwards
from $R_o$. } & 2.1 $\pm$ 0.7, 0.09 $\pm$ 0.03 \\
\hline
\end{tabular}
\end{minipage}
\end{table}

\subsubsection{The Nucleus}
\label{sec:nucleus}

As pointed out by \cite{gol94b}, 
there are disagreements in the location of the center of the galaxy, depending on how the
center is defined or which component is considered, 
consistent with a system that is highly disturbed.  
Fig.~\ref{fig:mom0_optical}, for example, shows a central minimum in the CO(J=3-2)
distribution  
(marked with a `C').  This minimum
agrees  with the minimum
seen in the CO(J=1-0) map \citep{gol94b,ran00a}
but is  15$^{\prime\prime}$ to the west of the
infrared (IR) center of Table~\ref{tab:galaxy_params} (marked with a star).

\begin{figure}
\includegraphics[scale=0.4]{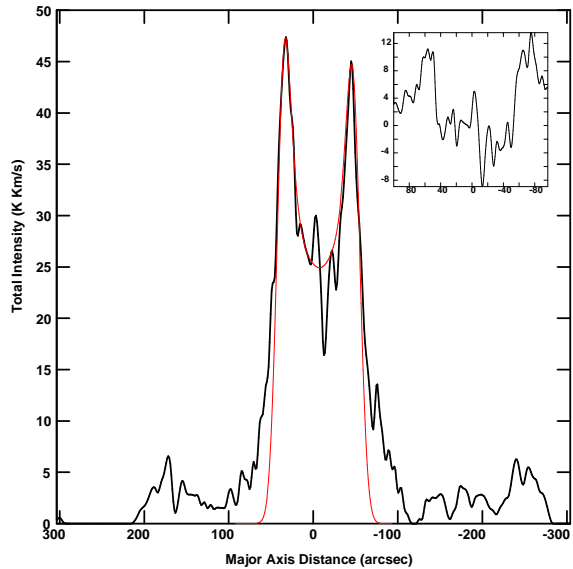}
\caption{Slice along the
 major axis of the total intensity CO(J=3-2) distribution shown in
Fig.~\ref{fig:mom0_optical}, averaged over a 
width 11$^{\prime\prime}$ (3 pixels) in $z$.
East is on the left and west is on the
right.  Offsets are with respect to the infrared center 
(Table~\ref{tab:galaxy_params}). The thin red curve shows a gaussian
ring with parameters as given in Table~\ref{tab:ring_model}. The inset
shows the residuals (data - model) for the central molecular ring.}
\label{fig:major_axis}
\end{figure}

The center of our modelled ring (Table~\ref{tab:ring_model}), however, 
agrees with the IR center within uncertainties, but not with the position of
C.  The IR center coincides with the central
radio peak \citep{gol99}, and our global CO(J=3-2) flux (not just the ring) is 
more symmetric
about the infrared center than about C.
  In Sect.~\ref{subsec:results:velocity} we will also present a 
dynamical argument for the nucleus to be more closely represented by the IR center.
 Therefore in this paper, 
when we refer to the center or nucleus, we mean the IR center of
Table~\ref{tab:galaxy_params}.  Note that there is a small {\it peak}
in CO(J=3-2) right at the nucleus as can be seen in 
Fig.~\ref{fig:major_axis} (see also Sect.~\ref{sec:velocity_ring}).



\subsubsection{The Weaker, Extended Disk Emission}
In addition to the strong central molecular ring, we also observe 
weaker CO(J=3-2) emission at much larger radii, evidently
 associated with the larger rotating galactic
disk (for example, see velocities of 499.3 km s$^{-1}$ and
738.3 km s$^{-1}$ in Fig.~\ref{fig:chanmapsa}).
This weaker
 broader disk emission appears distinct from the central molecular ring,
separated from it by the eastern minimum 
and western gap 
(see Fig.~\ref{fig:mom0_optical}); at these two locations,
there is also an abrupt change in the
rotation curve gradient (see Sect.~\ref{subsec:results:velocity}).
We will refer to this emission as the `outer disk' to distinguish it from
the central molecular ring.

This outer 
CO(J=3-2) disk tends to follow the optical
disk emission. 
 For example, the far eastern emission, centered
at a right ascension of about
 12$^{\rm h}$ 42$^{\rm m}$ 21$^{\rm s}$ (the `bump',
Fig.~\ref{fig:mom0_optical}) 
 maintains the bulging shape of the
underlying optical disk. 
On the far western side there is a distinct corrugation
in the CO(J=3-2) emission centered at a right ascension of
$\approx$  12$^{\rm h}$ 41$^{\rm m}$ 52$^{\rm s}$ (the `kink',
Fig.~\ref{fig:mom0_optical}). 
The outer disk also harbours the two expanding HI supershells found
by \cite{ran93} (marked with crosses in Fig.~\ref{fig:mom0_optical}).

The maximum extent of
the detected emission is 3.5$^\prime$ (9.25 kpc) to the east
of the nucleus and 4.7$^\prime$ (12.4 kpc) to the west.  This 
radial extent exceeds that of
previous CO observations in any transition, although the
CO(J=1-0) emission detected by \cite{gol94b} is almost as extensive
to the west.

\subsection{Comparison of Disk Emission with other Wavebands}
\label{sec:comparisons}

\begin{figure*}
\includegraphics[scale=0.7]{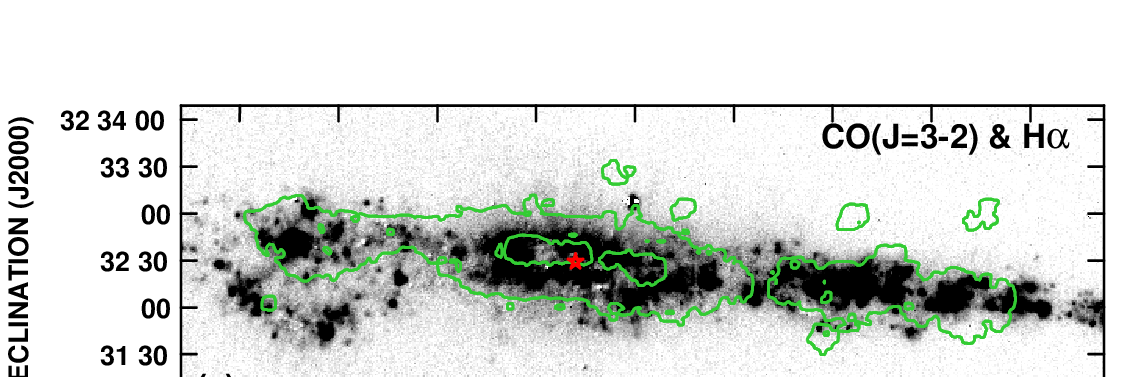}
\includegraphics[scale=0.7]{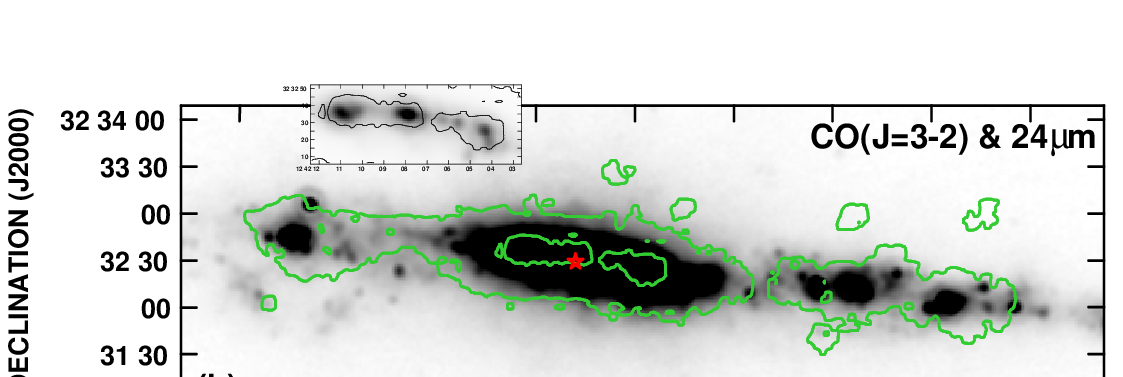}
\includegraphics[scale=0.7]{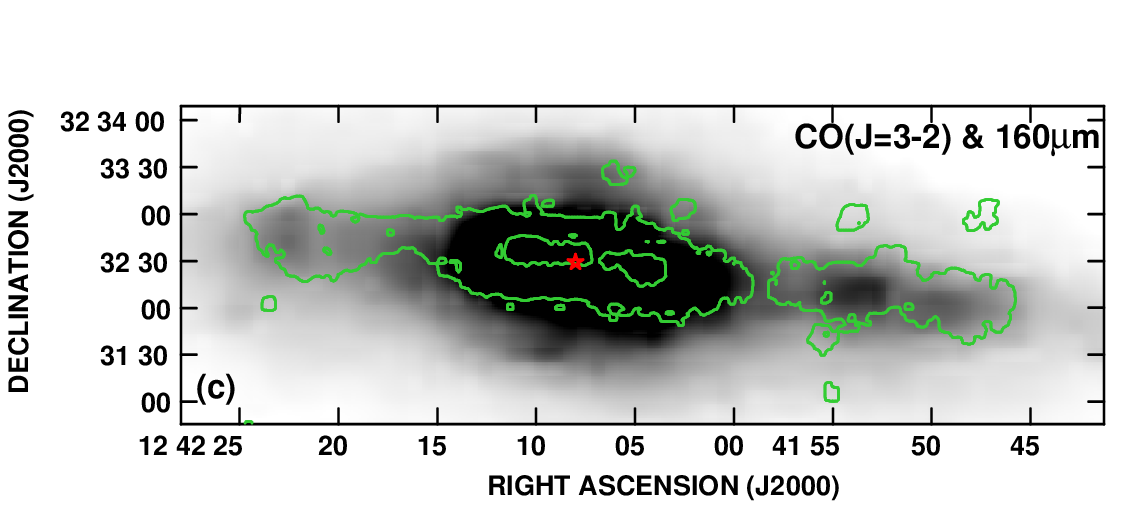}
\caption{Overlays of the lowest and second highest contours of
CO(J=3-2) emission from Fig.~\ref{fig:mom0_optical} on greyscales of
{\bf (a)} the H$\,\alpha$ image, {\bf (b)} the MIPS 24 $\mu$m image, and
{\bf (c)} the MIPS 160 $\mu$m image.  The greyscale images are shown at
their original resolutions which 
are 3.0$^{\prime\prime}$  for 
the H$\,\alpha$ image
  (from measurements of point sources
in the field),
6$^{\prime\prime}$ for the
24 $\mu$m image, and
 40$^{\prime\prime}$ for the 
160 $\mu$m image \citep{ben06}.
The greyscale ranges are arbitrary and have
been chosen to emphasize low to moderate intensity features. 
For the high contrast, high resolution 24 $\mu$m image, we also
show the central region in an inset.  The star denotes
the infrared center.}
\label{fig:overlays}
\end{figure*}

The CO(J=3-2) emission is displayed in a number of overlays
in Fig.~\ref{fig:overlays} which show distributions that
are taken to be proxies for unobscured
star formation (H$\alpha$), the hot dust distribution
(MIPS $\lambda\,$24 $\mu$m) and the 
cold dust distribution (MIPS $\lambda\,$160 $\mu$m).  For the high resolution,
high contrast $\lambda \,$24 $\mu$m map, we also show an inset of the central region.
It is well known that molecular gas, dust and star formation correlate
and we can see evidence for this in these overlays.  For example, the CO(J=3-2) bump
on the eastern side
is evident in the H$\alpha$ map as well as the $\lambda\,$24 $\mu$m
map, and the general trend of strong central emission and
weaker secondary
peaks at larger radii is common to the CO(J=3-2) and the two dust-emitting
bands.  

\begin{figure}
\includegraphics[scale=0.4]{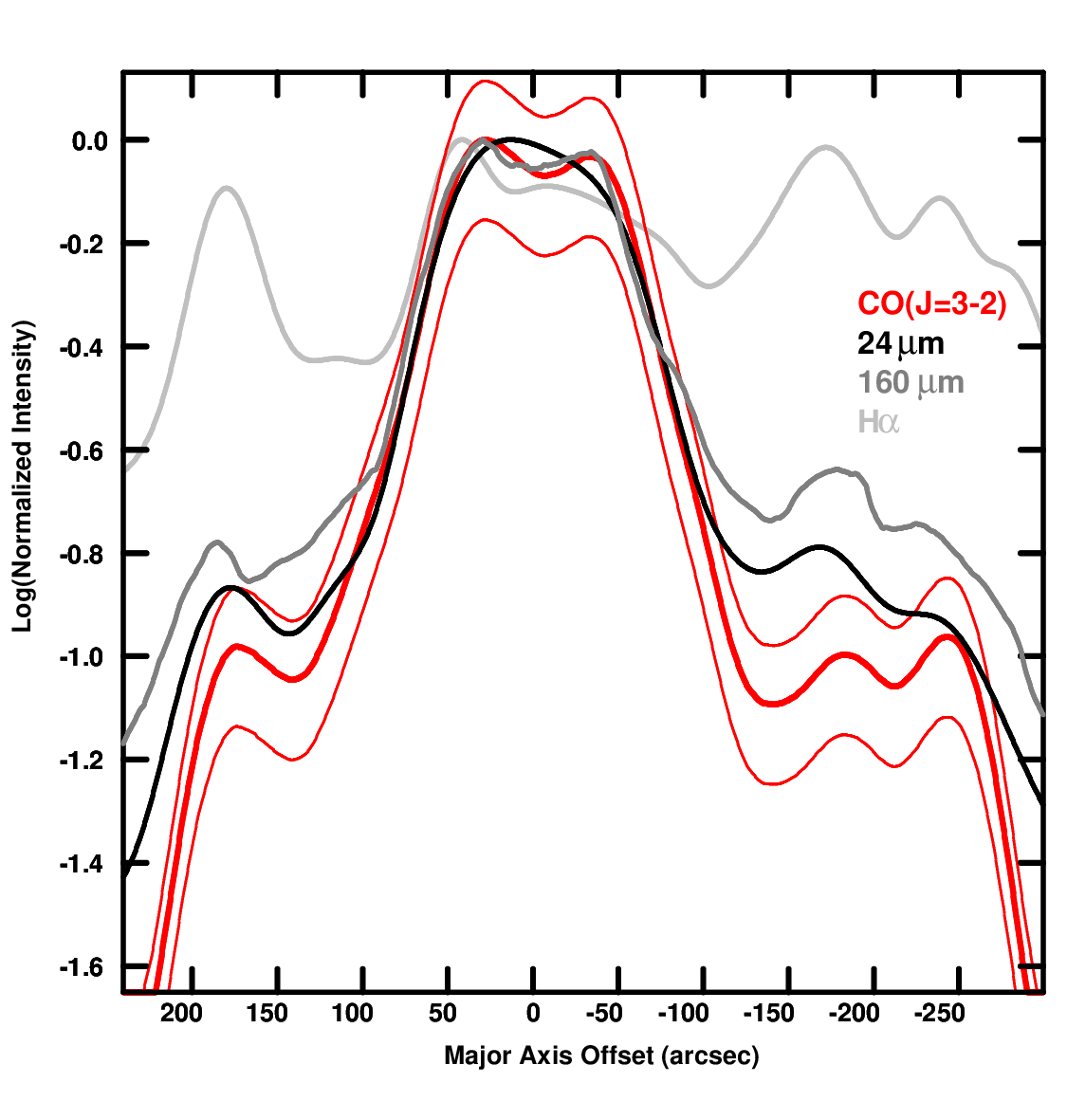}
\caption{Comparison of major axis slices at 40$^{\prime\prime}$
resolution, averaged over 40$^{\prime\prime}$ in $z$
and normalized to their peak values.  The positive x axis is to the east.
The thick red curve shows the CO(J=3-2) emission taken from Fig.~\ref{fig:mom0_optical}
and the thin red flanking curves show the extent of its error
bars (which are the dominant uncertainties in the figure). The steep fall-off
around +200 and -250 arcsec is due to the moment-generating routine.
The black curve is the $\lambda\,24$ $\mu$m emission from 
Fig.~\ref{fig:overlays}b, the medium grey curve represents the
$\lambda\,160$ $\mu$m emission from 
Fig.~\ref{fig:overlays}c and the light grey curve represents H$\alpha$
from Fig.~\ref{fig:overlays}a.
\label{fig:slices}}
\end{figure}

To explore these relations more quantitatively,
Fig.~\ref{fig:slices} further shows comparative major axis slices of 
CO(J=3-2),
$\lambda\,$24 $\mu$m, $\lambda\,$160 $\mu$m, and H$\alpha$ emission
at a common spatial resolution
\footnote{For the Spitzer images, resolutions were matched using
convolution kernels applicable to both the beam size and shape
\citep[see][]{gor08, ben10}.} and normalized  to their peak flux
values. Note that these slices contain most of
the emission in any of the maps, since the width of the slice
is 40 arcsec.  Here, the general trend of the two dust components following
the molecular gas distribution is again seen, but the departure of the H$\alpha$ emission
is more obvious since this latter component is strongly affected by
dust obscuration.   The correlation coefficient is 0.99 between CO(J=3-2) and
$\lambda\,$24 $\mu$m emission as well as between CO(J=3-2) and 
$\lambda\,$160 $\mu$m, whereas 
it is -0.001 between
CO(J=3-2) and H$\alpha$ emission.

Focussing only on molecular gas and dust,
Fig.~\ref{fig:slices} also shows that the CO(J=3-2) emission follows the 
 hot dust emission ($\lambda\,$24 $\mu$m) more closely than
the cold dust  ($\lambda\,$160 $\mu$m).  Colder dust
displays a broader distribution that declines more slowly from the central
 molecular ring. Between $\pm$ 200 arcsec, for example, the mean of the ratios, 
 CO(J=3-2)/$\lambda\,$24 $\mu$m
and CO(J=3-2)/$\lambda\,$160 $\mu$m along the major axis (not shown) 
are 0.91 and 0.35, respectively, supporting the closer connection between
CO(J=3-2) and hot dust.
This result is consistent
with the fact (see next section) that 
CO(J=3-2) is a good tracer of star formation and we would therefore
expect it to be more closely aligned with hotter dust.  Cold dust
is more likely to trace both molecular gas farther from star forming
regions as well as the
more broadly distributed atomic component.

\cite{ben06} have found that there is no significant variation with radius
between the $\lambda\,70$ $\mu$m,
$\lambda\,160$ $\mu$m, and $\lambda\,450$ $\mu$m emission in this galaxy
 and
so we would expect plots of $\lambda\,70$ $\mu$m and $\lambda\,450$ $\mu$m emission
to follow the cold dust emission of $\lambda\,160$ $\mu$m displayed in Fig.~\ref{fig:slices}.
 Available
 $\lambda\,450$ $\mu$m data do not
have sufficient signal-to-noise to test this but the close relation between
$\lambda\,70$ $\mu$m and $\lambda\,160$ $\mu$m data allows us to repeat
the above analysis 
at a higher spatial resolution (18$^{\prime\prime}$)
using $\lambda\,70$ $\mu$m as a cold dust indicator.
Again we find
the same conclusion that CO(J=3-2) more closely follows hot rather than 
cold dust.

Fig.~\ref{fig:slices} {\it appears} to show an exception to this result
right at the nucleus
where there is a peak in the $\lambda\,24$ $\mu$m hot dust distribution but minima in both 
CO(J=3-2) and $\lambda\,160$ $\mu$m cold dust.  However, 
at higher spatial resolution (see the inset of
Fig.~\ref{fig:overlays}b) we see that the nuclear peak at
$\lambda\,24$ $\mu$m is due to a strong `hot spot' located within
a region of approximately 17 arcsec (740 pc) diameter.  As pointed out
in Sect.~\ref{sec:nucleus}, there is also a small CO(J=3-2) peak at the nucleus
so the CO(J=3-2)/hot dust relation appears to hold even there, although relative 
emission strengths may vary.
We will show in the next section that the star formation
rate is higher at the nucleus than in the immediately surrounding region.




\subsection{CO(J=3-2) Excitation}
\label{sec:results:excitation}

\begin{figure}
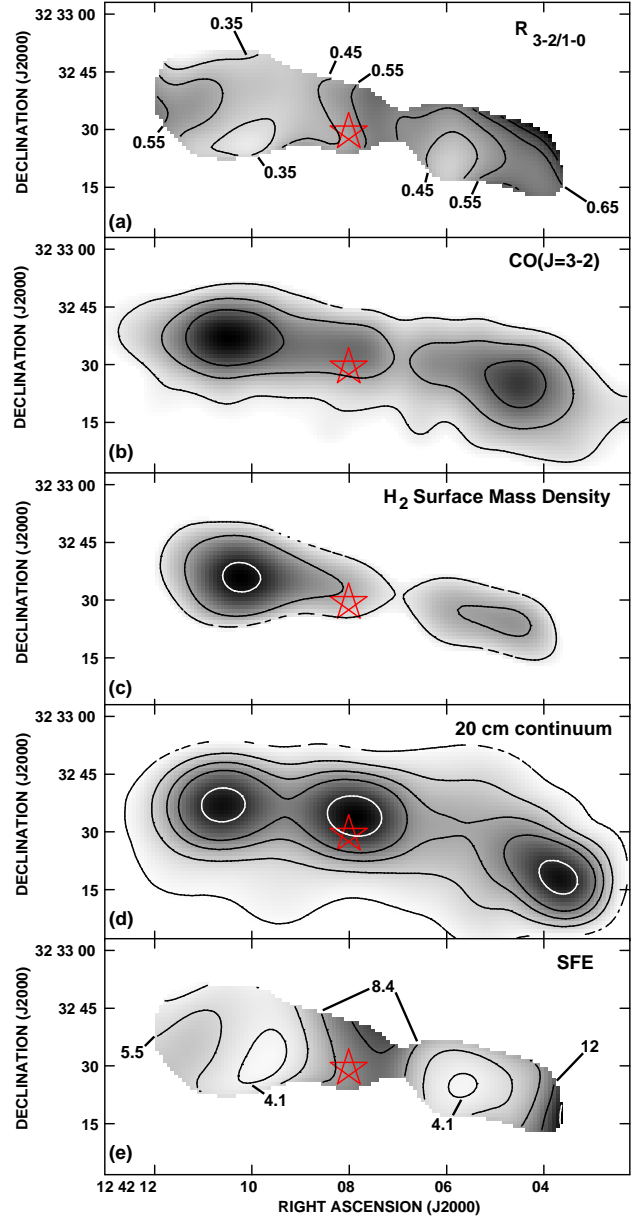

\includegraphics[scale=0.45]{CO_ratio_new.ps}
\includegraphics[scale=0.45]{CO_32_smoth_new.ps}
\includegraphics[scale=0.45]{mass_density_new.ps}
\includegraphics[scale=0.45]{first_20cm_new.ps}
\includegraphics[scale=0.45]{sfe_new.ps}
\caption{Comparison of emission in the brightest part of the
central molecular ring.  All maps have been smoothed to 
17$^{\prime\prime}$ resolution, cut off at the
5$\sigma$ level, and are represented in both contours
and greyscale with darker representing higher values.
The star marks the galaxy's center.
{\bf (a)} The integrated intensity ratio map, $R_{3-2/1-0}$.
Contours are at 
0.35, 0.45, 0.55, 0.65, and 0.75.
{\bf (b)} The JCMT CO(J=3-2) map in greyscale and contours
of the inner region of 
Fig.~\ref{fig:mom0_optical}.
Contours are at 11, 22, and 33 K km s$^{-1}$ and the peak is
48.9 K km s$^{-1}$.
{\bf (c)} Molecular hydrogen 
surface mass density, $\Sigma_{H_2}$,
 formed from the CO(J=1-0) total intensity
map, adopting X = $2.0\,\times\,10^{20}$ mol cm$^{-2}$ (K km s$^{-1}$)$^{-1}$.
 Contours are at 125, 200, and 350 M$_\odot$ pc$^{-2}$ and the
peak is 380 M$_\odot$ pc$^{-2}$.
{\bf (d)} Radio continuum emission from the FIRST survey. Contours are
at 3, 6, 9, 12 and 18 mJy beam$^{-1}$ and the peak is at
21.4 mJy beam$^{-1}$. The transformation to $\Sigma_{SFR}$ is given
in Eqn.~\ref{equation3}.
{\bf (e)} $SFE$, with contours at
4.1, 5.5, 8.4, and 12 $\times\,10^{-10}$ yr$^{-1}$.  The peak is 17.3
$\times\,10^{-10}$ yr$^{-1}$.}
\label{fig:ratio_mass}
\end{figure}

We can study the CO(J=3-2) excitation in NGC~4631 by forming a
ratio map of
 CO(J=3-2) to CO(J=1-0) emission.  To this end,
we use CO(J=1-0) data obtained using the 
Berkeley Illinois Maryland Array (BIMA), originally 
 at resolution of $\approx$ 8$^{\prime\prime}$,
kindly supplied by R. J. Rand \citep{ran00a}.  The total intensity
CO(1-0) map from \cite{ran00a} (his figure 1) was interpolated to the same
grid as our total intensity CO(J=3-2) map and then both images were smoothed
to 17$^{\prime\prime}$ resolution to reduce
noise in the ratio map.  
The CO(J=1-0) map was then converted\footnote{The Planck relation
was used. In the Rayleigh-Jeans limit,
the conversion, for a beam solid angle of $\Omega_b\,=\,
7.7\,\times\,10^{-9}$ sr, is 
1 Jy beam$^{-1}$ km s$^{-1}$ = 3.1 K km s$^{-1}$.}
 to units of 
K km s$^{-1}$.
As a check on the CO(J=1-0) map, we made comparisons
with previous single dish  
 CO(J=1-0) data from the literature.
%
As indicated by \cite{ran00a}, the
BIMA total flux agrees with that of \cite{gol94b} within uncertainties.  Also,
the BIMA integrated intensity agrees with the 
Institut de Radioastronomie Millim{\'e}trique (IRAM)
value
 at the same central position and resolution as given in
\cite{isr09}\footnote{The BIMA result is
41 K km s$^{-1}$ compared to the the IRAM result of 45 $\pm$ 6
K km s$^{-1}$.}.  A comparison of major axis slices between the BIMA data and
IRAM data of \cite{gol94b} at the same resolution also shows good agreement
in the region of the central molecular ring.
Both the BIMA map and our JCMT CO(3-2) map
at 17$^{\prime\prime}$ resolution were
then cut off at a conservative 5$\sigma$ level before forming the
ratio. 
The
 resulting CO(J=3-2)/CO(J=1-0) map, $R_{3-2/1-0}$, 
shown in Fig.~\ref{fig:ratio_mass}a,
 could be formed only over
the brightest, inner 1.8$^{\prime}$ diameter (4.7 kpc) part of the central
molecular ring, i.e. approximately over the FWHM of the emission shown
in Fig.~\ref{fig:major_axis}. The matching smoothed JCMT CO(J=3-2) map is
shown in Fig.~\ref{fig:ratio_mass}b.

To estimate the uncertainty in the $R_{3-2/1-0}$  map, 
we have formed a relative error
map (not shown), 
based on random errors in the two individual maps. The average of the
absolute values of the relative errors, is 11\% (lower in regions of
higher S/N).  Errors due to positional offsets are much lower than
this.
 Allowing for an additional
absolute calibration error due to uncertainties in $\eta_{MB}$ 
(Sect.~\ref{sec:obs_datared}), we estimate that the average
uncertainty in Fig.~\ref{fig:ratio_mass}a is
of order 25\%.  However, the calibration error can be neglected when
examining the distribution of $R_{3-2/1-0}$ over the map since it
would shift all points the same way.
We have verified that our value of 
$R_{3-2/1-0}$ agrees with the result of 
\cite{isr09} for the central position when smoothed to the same resolution.
 

The average value of 
$R_{3-2/1-0}$ over the region displayed in Fig.~\ref{fig:ratio_mass}a
is 0.47 with an rms of 0.11 and extrema of 0.24 and 0.92.  Ninety-five
percent of all pixels lie within the range, 0.28 to 0.66.
These variations in $R_{3-2/1-0}$ appear to be real since they
exceed the
approximately 11\% uncertainty discussed above.

Lower values of $R_{3-2/1-0}$, on average, are generally found for
galaxy disks or for galaxies globally, for example those that are
typically found in the molecular medium along the Milky Way disk
\citep[$\approx$ 0.4,][]{san93}, the disks of galaxies,
or samples of nearby galaxies \citep[0.2 to 0.7,][]{mau99,wil09}.
Higher values, on the other hand, are seen 
 in the central regions 
of galaxies \citep[e.g. 0.7 to 1.2 for M~83, 0.9 for M~82, 0.6
for M~51,][respectively]{mur07,til91,isr06}, in
high excitation regions or regions associated with star formation \citep[e.g.
up to 1.6 for the Antennae and up to 1.2 for specific regions
in M~33,][respectively]{pet05, tos07}, or in galaxies in which temperatures
and/or molecular gas densities are higher, 
on average \citep{mau99}.  Therefore, the line ratios found in the
central molecular ring of NGC~4631 appear to be typical of galaxy
disks rather than of regions containing strong starbursts.

\begin{figure}
\includegraphics[scale=0.34]{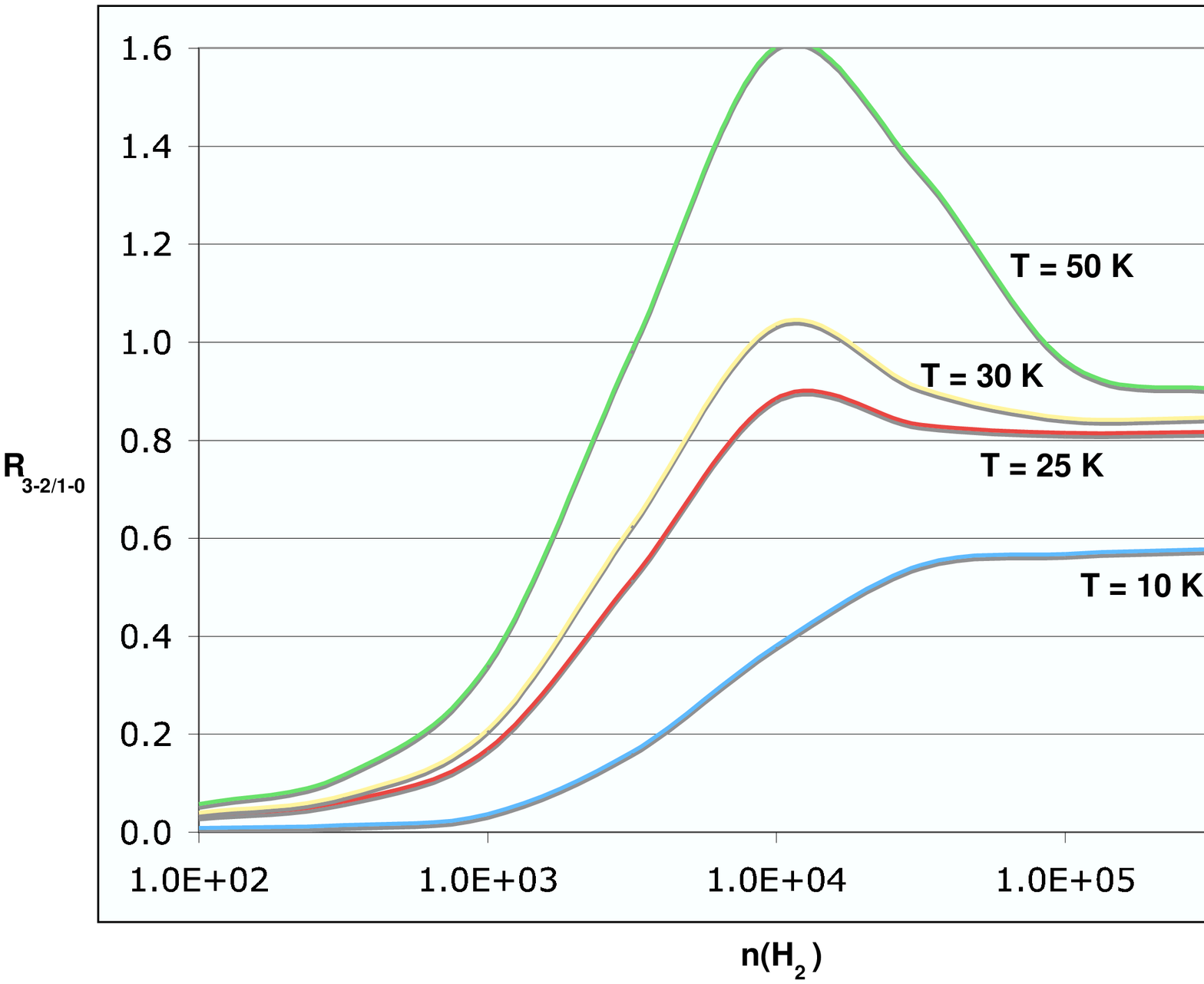}
\caption{Plot of the ratio, $R_{3-2/1-0}$ as a function of molecular hydrogen
density, $n_{H_2}$ (cm$^{-3}$) for a range of temperature (labelled)
from an LVG code, assuming an abundance per unit velocity gradient of
 $X_{^{12}CO}/(dv/dR)\,=\,10^{-6}$ pc  (km s$^{-1})^{-1}$ and a single component model.}
\label{fig:LVGplot}
\end{figure}

For the central 21 arcsec diameter region of NGC~4631 only,
\cite{isr09} 
finds that the molecular gas
 is best
represented by two molecular components, with most of the mass (80\%) 
in a cold ($T_{kin}\,=\,10$ K), tenuous (n$_{H_2}\,
\approx\,300$ cm$^{-3}$) component, and a lesser amount (20\%) in a
warm ($T_{kin}\,=\,150$ K), moderately dense  (n$_{H_2}\,
\approx\,10^3$ cm$^{-3}$)
state. Fig.~\ref{fig:ratio_mass}a has now extended the measurement of
$R_{3-2/1-0}$ over a much larger region than that measured by \cite{isr09}.
With a 
single line ratio, we cannot place strong constraints on the state of
the molecular gas over the extended region.
In general, however, if a {\it single} physical state is
present, the relationship between kinetic temperature,
molecular gas density, and $R_{3-2/1-0}$ can be represented by
 Fig.~\ref{fig:LVGplot} in which we show
a large velocity gradient (LVG) model 
 \citep[e.g.][]{gol74,irw92} with an
 abundance
per unit velocity gradient of $X_{^{12}CO}/(dv/dR)\,=\,10^{-6}$
pc  (km s$^{-1})^{-1}$ \citep[the latter value
 as suggested by the results of][]{zhu03}.
Values of $R_{3-2/1-0}$ from 0.28 to 0.66 would then
imply gas densities of order $10^3$ cm$^{-3}$ over the range of
temperatures shown\footnote{If $X_{^{12}CO}/(dv/dR)$ changes by an 
order of magnitude
\citep[see][]{zhu03}, then over the range of
interest, the resulting density changes by less than a factor of two.}.
Note that this density refers
to the individual cloud densities
that are responsible for CO excitation, rather
than mean densities within the beam.
If {\it two} components are present throughout the region 
with a lower density component dominating, such as found by 
\cite{isr09} for
the center, then $10^3$ cm$^{-3}$
should be an upper limit to $n_{H_2}$.  Therefore, again we find that the
conditions within
molecular clouds in the central region of NGC~4631 appear to be typical
of low density molecular gas regions in galaxy disks
(i.e.  $n_{H_2}\,<\,10^4$ cm$^{-3}$)
 rather than the $\gtabouteq\,10^4$
cm$^{-3}$ which are more typical of central starburst regions
\citep[see, e.g.][]{wei01, hai08, ion07}.

\subsection{Molecular Mass and Gas/Dust Ratio}
\label{sec:results:mass}

The distribution of molecular hydrogen mass 
surface density, $\Sigma_{H_2}$, in the central molecular ring of
NGC~4631 is shown in Fig.~\ref{fig:ratio_mass}c.  This map was formed
from the \cite{ran00a} CO(J=1-0) map
and therefore shows essentially the CO(J=1-0) distribution. We have used
a standard and constant
CO(J=1-0) integrated intensity
 to H$_2$ conversion factor of
X = $2.0\,\times\,10^{20}$ mol cm$^{-2}$ (K km s$^{-1}$)$^{-1}$,
consistent with \cite{wil09},
corresponding to 
3.2 M$_\odot$ pc$^{-2}$ (K km s$^{-1}$)$^{-1}$ of molecular hydrogen
(uncorrected for heavier elements). The mass distribution shows similar peaks and
minima as the
CO(J=3-2) map; differences are highlighted by the ratio map 
 Fig.~\ref{fig:ratio_mass}a.
For this value of X the total
 molecular hydrogen mass over the region shown in Fig.~\ref{fig:ratio_mass}c is
M$_{H_2}\,=\,7.4\,\times\,10^8$ M$_\odot$.

For this galaxy, there have
been several measurements of the value of X from independent line ratio
analyses.  \cite{isr09} finds X = $0.3\,\times\,10^{20}$ mol 
cm$^{-2}$ (K km s$^{-1}$)$^{-1}$ within the central 21 arcsec and
\cite{pag01} finds X = $0.5^{+2.0}_{-0.3}\,\times\,10^{20}$ mol 
cm$^{-2}$ (K km s$^{-1}$)$^{-1}$ within the central 46 arcsec and
 X = $1.8^{+7.1}_{-1.0}\,\times\,10^{20}$ mol 
cm$^{-2}$ (K km s$^{-1}$)$^{-1}$
outside of the central 46 arcsec.  The two \cite{pag01} values agree
with each other and also with our adopted value within uncertainties.
If we adopt the \cite{isr09} value of X for the central 21 arcsec only, then 
M$_{H_2}$ reduces  
by approximately 11\% which is not significantly different from the above
result,
given the uncertainties.  Although adopting a lower central value of
X does not
significantly perturb our calculation of total mass obtained from
Fig.~\ref{fig:ratio_mass}c, there would be changes in the appearance
of this map, should X vary with position.
 
In addition, since we have
observed CO(J=3-2) over a larger region than
shown in Fig.~\ref{fig:ratio_mass}, and to larger radii
than previously detected in any CO transition, we
can estimate the total molecular hydrogen mass
in NGC~4631 by applying an appropriate
value of $R_{3-2/1-0}$ over the entire emission shown in
Fig.~\ref{fig:mom0_optical} and using 
the standard value of X listed above.   Adopting the mean value from the central
region,
$R_{3-2/1-0}$ = 0.47, we find a total mass of
  M$_{H_2}\,=\,2.2\,\times\,10^9$ M$_\odot$
with an error of $\approx$
25\% which represents the uncertainty in Fig.~\ref{fig:ratio_mass}a 
including the calibration error.  This uncertainty does not include
uncertainties in X or in its possible variation with position in the galaxy.
Our result
has improved on that 
of M$_{H_2}\,=\,1.5\,\times\,10^9$ M$_\odot$ (adjusted to our distance
and value of X) provided by \cite{gol94a},
 whose map does not extend as far out as ours
and who suggested
a factor of 2 uncertainty on their quantity.

The total molecular gas mass is
 22\% of the total HI mass of
M$_{HI}$ = 1.0 $\times\,10^{10}$ M$_\odot$ found by \cite{ran94}
(Table~\ref{tab:galaxy_params}).
The total HI + H$_2$ mass is therefore M$_{HI+H_2}\,=\,1.22\,\times\,10^{10}$ M$_\odot$ and
dominated by HI.  
Adjusted for heavy elements (a factor of 1.36), the total gas mass
is then
 M$_{g}$ = $1.66\,\times\,10^{10}$ M$_\odot$
The total dust mass in NGC~4631 is estimated to be
M$_d\,=\,9.7\,\times\,10^7$ M$_\odot$ \citep{ben06}, leading to a global
gas-to-dust ratio of 170, a value that is typical of spiral galaxies, including
the Milky Way \citep[e.g.][]{dra07}.
These masses are integrated over the entire galaxy and do not necessarily represent
the relationships between atomic, molecular, and dust components in individual regions; we do
not have sufficient information to determine region-specific quantities without
a model for each of those components in this edge-on galaxy.

\subsection{Star Formation}
\label{sec:results:star_formation}

In Table~\ref{tab:galaxy_params}, we list two
 estimates of the global SFR, the first from the FIR luminosity, and the second from
the H$\,\alpha$
emission corrected for extinction using 
$\lambda\,24$ $\mu$m 
data (H$\,\alpha_{corr}$).
 The two values differ by
about a factor of two.  SFRs can be determined from a variety of different tracers
but in a galaxy as edge-on as NGC~4631, optical depth effects can become large, uncertain, and
can vary in an irregular fashion with position.
The empirical relation for determining H$\,\alpha_{corr}$, for
example, although considered relatively robust, has not been determined for galaxies that
are edge-on \citep{cal07}.  
Fig~\ref{fig:slices} also confirms that there are large differences between
the shape of the 
H$\,\alpha$ curve and other tracers that are not as badly affected by
extinction.  To probe the spatially resolved SFR, then, H$\,\alpha_{corr}$ 
cannot be used with full confidence (see also Footnote~\ref{sfr}) and the FIR luminosity does not have sufficient
resolution\footnote{The FIR data have spatial resolutions of 1.44 arcmin and 2.94 arcmin at
$\lambda\,60$ $\mu$m and $\lambda\,100$ $\mu$m, respectively \citep{san03}.}.

An alternative is to use radio continuum emission which requires no correction for
optical depth effects and for which we have data
for the high signal-to-noise region at the center of the galaxy
 as shown in
Fig.~\ref{fig:ratio_mass}d, taken from the
Faint Images of the Radio Sky at Twenty-Centimeters (FIRST) survey 
\citep{bec95} and smoothed from an original spatial resolution of 5 arcsec.
The FIRST survey is insensitive to structure greater than
2$^\prime$ in scale which is approximately the diameter of the region
shown in Fig.~\ref{fig:ratio_mass} and therefore there should be
no missing flux on the scales that we are probing. 
In addition, although CR electrons diffuse from their source of origin
in comparison to other tracers of SF, the scale over which this
occurs for NGC~4631 \citep{mar98}
is less than the beam size of Fig.~\ref{fig:ratio_mass}.
Finally, there is no evidence for a radio emitting AGN (or candidate)
in this galaxy at
a flux level that could affect the SFR determination 
\citep[see][]{gol99}.  
The radio continuum is therefore a good measure
of massive SF in the displayed region from which we can then estimate the total
SFR\footnote{A map of H$\alpha_{corr}$ 
should resemble Fig.~\ref{fig:ratio_mass}d if the former has been adequately corrected
for extinction.  We have verified that there
are sufficient differences between the two maps
and therefore
Fig.~\ref{fig:ratio_mass}d
is the preferred tracer of spatially resolved SF.
\label{sfr}}. 

We first use the
relation given in \cite{con92} which provides only the {\it massive}
($M\,>\,5~M_\odot$)
SFR, SFR$_{m}$ from the radio emission, i.e.  
\begin{equation}
\frac{{\rm SFR}_{m}}{[{\rm M}_\odot~ {\rm yr}^{-1}]} \,
=\,
2.47\,\times\,10^{-22} \,\frac{L_{1.4}}{[{\rm W}~ {\rm Hz}^{-1}]}
\label{equation1}
\end{equation}
where  
$ L_{1.4}$ is the 1.4 GHz radio luminosity per unit bandwidth and
we have assumed that the 1.4 GHz emission is dominated by the
non-thermal component\footnote{\cite{gol99} estimates a thermal fraction of
10\% at 5 GHz, indicating that the fraction will be lower at
1.4 GHz.} with a typical
spectral index of 0.8.

 Integrated over the region displayed in Fig.~\ref{fig:ratio_mass}d, we find
SFR$_{m}$ = 0.34  $M_\odot$ yr$^{-1}$ (or $\nu\,=\,0.014$ SNe yr$^{-1}$, from
relations in Condon, 1992).   
We can also obtain SFR$_m$ for the entire galaxy (not just the region
shown in Fig.~\ref{fig:ratio_mass})
from Eqn.~\ref{equation1} 
 and the global radio continuum flux 
\citep[771.7 mJy][]{str04b} resulting in SFR$_{m}$ = 1.8
${\rm M}_\odot$ yr$^{-1}$.  Therefore, $\approx\,$19\% of the massive SF
is occurring in the region shown in Fig.~\ref{fig:ratio_mass}.  
Thus, the massive star formation in this galaxy is widely distributed
in contrast to `nuclear starbursts' such as M~82
\citep[see also arguments in][]{tul06b}.  The   
Galaxy Evolution Explorer (GALEX) UV image also reveals widely
distributed UV emission consistent with distributed star formation in
this galaxy \citep{dep07}.


We can now apply a
factor to account for non-massive star formation, i.e. 
SFR/SFR$_m\,=\,2.4$, where SFR is taken to be
SFR$_{FIR}$ from
Table~\ref{tab:galaxy_params} and SFR$_m$ is given above. Note that this 
correction factor is 
 mid-range between the correction factors that would result by applying
a disk IMF given by  \cite{cha03} (a factor of 1.1) and a Salpeter IMF
(a factor of 5.5) over a total mass range of
$0.1\,\le\,{\rm M}\,\le\,100\,{\rm M}_\odot$ yr$^{-1}$.
The result, which converts the values of Fig.~\ref{fig:ratio_mass}d to SFR per unit
area, is

\begin{equation}
\frac{\Sigma_{SFR}}{[{\rm M}_\odot~ {\rm yr}^{-1}\,{\rm pc}^{-2}]}
\,=\,
9.2\,\times\,10^{-6}\,
\frac{I_{1.4}}{[{\rm Jy}~ {\rm beam}^{-1}]}
\label{equation3}
\end{equation}

We now provide a measure of the efficiency
of star formation, $SFE$,
in the central regions of NGC~4631 by forming the ratio map of
${\rm SFE}/[{\rm yr}^{-1}]\,=\,\Sigma_{\rm SFR}/[{\rm M}_\odot~ {\rm yr}^{-1}~{\rm pc}^{-2}]
/\Sigma_{H_2}/[{\rm M}_\odot~{\rm pc}^{-2}]$, which assumes that the
stars mainly form in molecular gas (excluding HI).  The result is shown
in Fig.~\ref{fig:ratio_mass}e.
Note that we make no correction for inclination, so the values in
this figure represent SFEs over the line of sight through the region.
The mean value is ${\rm SFE}\,=\,6.4\,\times\,10^{-10}$ yr$^{-1}$ with an
uncertainty of $2.5\,\times\,10^{-10}$ yr$^{-1}$
(based on the rms of the map) and extrema of 
$3.5$ and $17.3\,\times\,10^{-10}$ yr$^{-1}$. 
This result falls within the 1$\sigma$ error bar of the sample
of \cite{row99} (after conversion to their definition of SFE\footnote{The
transformation given in \cite{ken98} was used to obtain an H$\alpha$ SFR but
no inclination correction is made.}) who
examined global SFEs for 568 galaxies; that is, although there is a strong
concentration of molecular gas in the central region of NGC~4631 and the
galaxy is interacting (Sect.~\ref{sec:intro}),
the SFE in this
region is typical of galaxy disks in general.

The inverse of SFE is a simple measure of the gas consumption timescale,
$t_g$,
if the SFR remains constant with time.  Following \cite{kna09} and references
therein, including a correction for recycling of material by stars into the ISM
yields
$t_g\,=\,1/(0.6\,{\rm SFE})$.  
  For
the region shown in Fig.~\ref{fig:ratio_mass}e, we find a mean timescale of
$\approx\,2.6\,\times\,10^{9}$ yr (to within approximately a factor of two, given the
above uncertainties).
This result will be a lower limit
to the total gas consumption timescale since it does not include HI.  
From the arguments of the previous section, we expect the mean
value of $t_g$ to increase by a
factor of 2 or 3, should HI be included.  All uncertainties considered, this
result  
is still consistent with values found in other galaxies
\citep{kna09, big08, ler08, ken94, gol94b}. 
Clearly, the central region of NGC~4631 contains a strong build-up
of molecular gas (Fig.~\ref{fig:major_axis}), but the gas consumption
timescale is long for a constant SFR.  We return
to this point in Sect.~\ref{sec:discussion}.

There is clearly a similarity between the $R_{3-2/1-0}$ map
(Fig.~\ref{fig:ratio_mass}a), representing 
molecular gas excitation (higher density and/or higher temperature as
indicated by
Fig.~\ref{fig:LVGplot})
and the SFE map (Fig.~\ref{fig:ratio_mass}e) representing the
star formation rate per unit molecular
gas mass.  Regions of lower ratio, as noted in
Sect.~\ref{sec:results:mass} approximately correspond to regions of
lower SFE -- a trend also observed in M~83 by \cite{mur07}.
There
are still differences, however.  For example, a map of the ratio of
SFE/$R_{3-2/1-0}$ (not shown) 
results in an rms variation of 25\%, the 
most important difference being
at the nucleus at which
the SFE appears to be enhanced in comparison to $R_{3-2/1-0}$.  
Since
both of these maps have been formed by normalization with
 the CO(J=1-0) distribution, the nuclear enhancement can be easily seen
by directly comparing the CO(J=3-2) distribution of Fig.~\ref{fig:ratio_mass}b
(from which $R_{3-2/1-0}$ has been formed) to the
20 cm radio continuum map of Fig.~\ref{fig:ratio_mass}d (from which SFE
was derived).   The radio continuum map shows a strong nuclear peak whereas
the CO(3-2) map does not.  Thus, there is an enhancement in SFR and SFE right at the
nucleus in comparison to the surrounding region.

\subsection{The Velocity Distribution}
\label{subsec:results:velocity}

\begin{figure}
\includegraphics[scale=0.35]{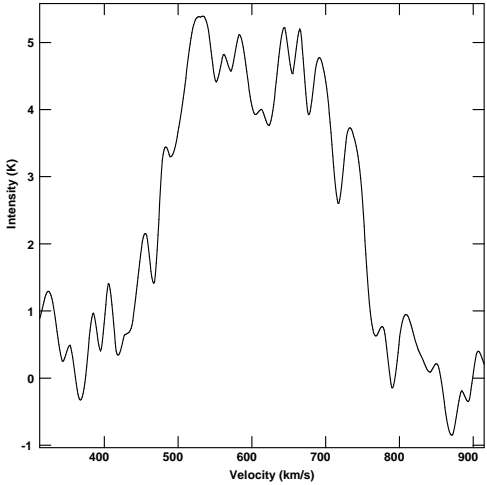}
\caption{Global profile formed from the 3 central frames of Fig.~\ref{fig:pv}.  The rms is
0.5 K. 
The
velocity is heliocentric.}
\label{fig:global_profile}
\end{figure}

\begin{figure}
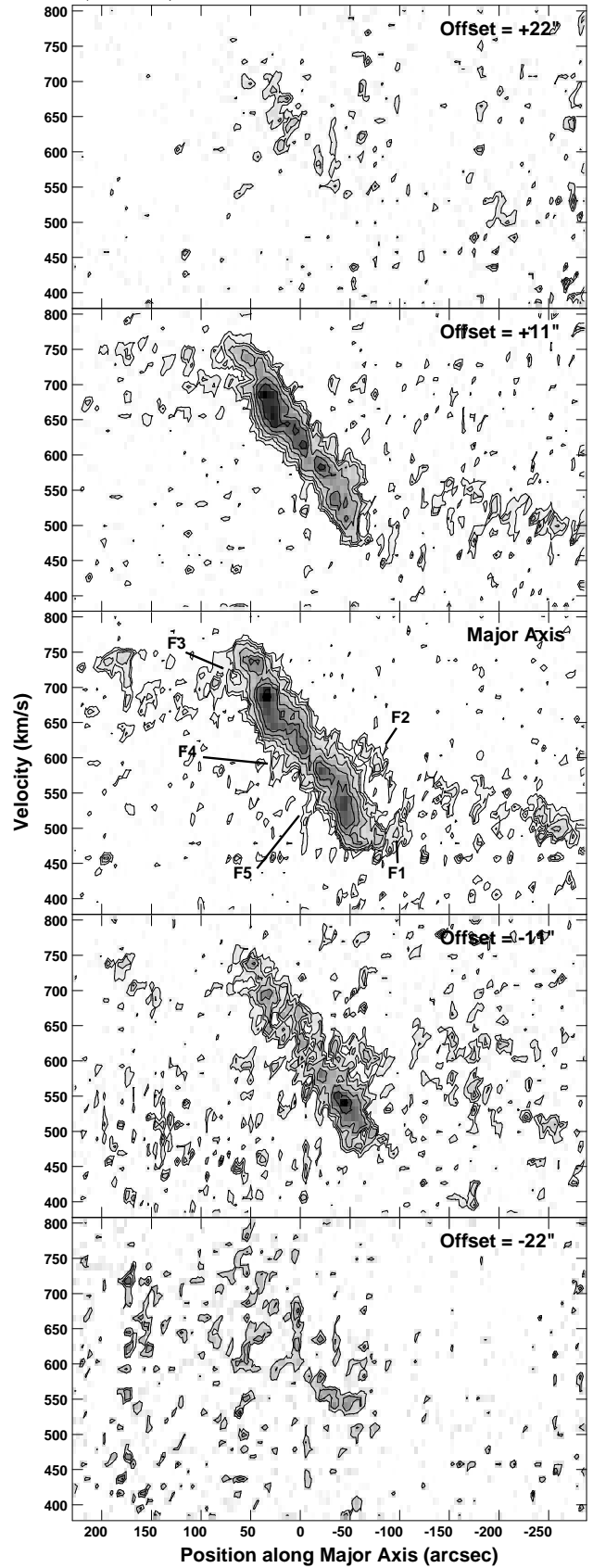

\includegraphics[scale=0.44]{pv++.ps}
\includegraphics[scale=0.44]{pv+.ps}
\includegraphics[scale=0.44]{pv.ps}
\includegraphics[scale=0.44]{pv-.ps}
\includegraphics[scale=0.44]{pv--.ps}
\caption{PV plots parallel to the major axis averaged
over a width of 11$^{\prime\prime}$ (contours on greyscale). 
Vertical offsets
(positive to the north) are marked on each
frame.
Contours are at 0.044 (2$\sigma$), 0.07, 0.10, 0.16, 0.25, 0.35, and 0.6
K. Positive is to the east on the x axis
(50$^{\prime\prime}$ = 2.2 kpc).  
F1-F5 are discussed
in Sect.~\ref{subsec:results:supershells}.
}
\label{fig:pv}
\end{figure}

The global profile and position-velocity (PV) slices along and parallel
to the major axis of NGC~4631 
are shown in
Fig.~\ref{fig:global_profile} and Fig.~\ref{fig:pv}, respectively.

It is unusual to show a CO(J=3-2) global profile for a galaxy and 
Fig.~\ref{fig:global_profile} serves to illustrate the high quality of
the HARP data.  From this plot, 
   the line width at
half-maximum is $W_{50}\,=\,270\,\pm\,10$ km s$^{-1}$ and at 
20\% of maximum is  $W_{20}\,=\,316\,\pm\,10$ km s$^{-1}$.  From the
mean midpoint of the linewidths, we find $V_{sys}\,=\,607\,\pm\,10$ km s$^{-1}$,
in good agreement with the HI value of Table~\ref{tab:galaxy_params}
as well as the HI value found by \cite{ran94} (610 km s$^{-1}$) though it is
somewhat lower than the CO(J=1-0) value of \cite{gol94b}
(628 km s$^{-1}$) and \cite{gao04} (658 km s$^{-1}$)\footnote{All values corrected to
our velocity definition, where necessary.}.  
This global CO(J=3-2) value
of $V_{sys}$ is plotted in Fig.~\ref{fig:features}a.



The PV velocity distribution shown in Fig.~\ref{fig:pv} 
will be considered in three
parts, namely,
the central molecular ring and nucleus, 
the outer disk emission (both discussed below) and anomalous
features and high latitude gas (to be discussed in Sect.~\ref{subsec:results:faint}).

\subsubsection{The Central Molecular Ring and Nucleus}
\label{sec:velocity_ring}

The most dominant emission shown in Fig.~\ref{fig:pv}
is, again, the central molecular ring
which extends 3.8$^{\prime}$ (10 kpc) in diameter (as noted in
Sect.~\ref{sec:results:distribution})
forming a strongly emitting region with a steeply rising rotation curve.
The velocity gradient is
2.1 km s$^{-1}$ arcsec$^{-1}$, or 48 km s$^{-1}$ kpc$^{-1}$
with an estimated 25\% uncertainty depending on where the slope is measured.
The fact that the western peak is stronger to the south (offset of -11$^{\prime\prime}$)
and the eastern peak is stronger to the north (offset of +11$^{\prime\prime}$) reflects
a slight asymmetry in the curvature of the major axis. (The major axis position angle
of 6.5 deg 
was adopted to pass through both eastern and western maxima.)  These gradients agree
with those of the CO(J=1-0) distribution \citep{ran00a}.

\begin{figure*}
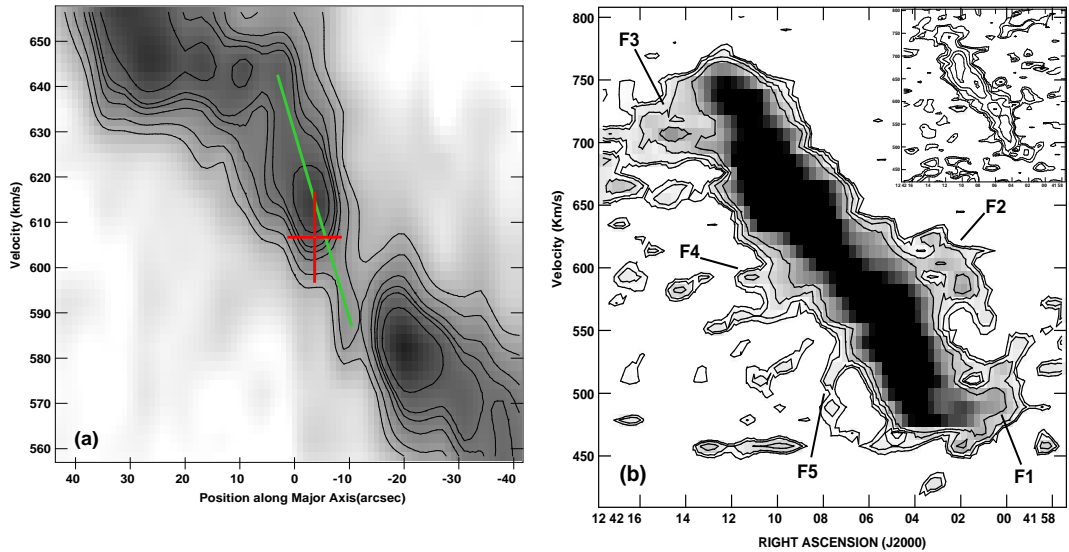

\includegraphics[scale=0.38, width=2.8truein, height=3.1truein]{core.ps}
\includegraphics[scale=0.4,width=2.8truein,height=3.1truein]{pvfeatures.ps}
\caption{Enhanced versions of the major axis frame of Fig.~\ref{fig:pv}.  {\bf (a)}
Blow-up of the nuclear region emphasing the brightest emission with contours at 
0.2, 0.25, 0.3, 0.33, 0.35 and 0.4 K.
The x axis is centered on the IR center (Table~\ref{tab:galaxy_params}). 
The cross
marks the location of the modelled ring center in x
(Table~\ref{tab:ring_model}) and $V_{sys}$ (from the total global profile of
Fig.~\ref{fig:global_profile}), with the cross size
delineating their respective error bars.  The green line shows the nuclear gradient
(see Sect.~\ref{sec:velocity_ring}).
{\bf (b)}  Here, the major axis has been smoothed spatially with a Gaussian of
FWHM = 11$^{\prime\prime}$ to emphasize the fainter emission.
 Contours are at 0.028 ($2\,\sigma$), 0.035, 0.050 and 0.074 K.
Features have been labelled as in Fig.~\ref{fig:pv}.  {\it Inset:} CO(J=1-0) major axis emission
from the original data of Rand (2000a) rotated and smoothed as in the main figure; contours have
been arbitrarily set.}
\label{fig:features}
\end{figure*}


An interesting new result is the
`kink' in the rotation curve approximately at the infrared
nucleus of NGC~4631 
such that
in a single pixel (0.16 kpc) there is a vertical drop in velocity
 (east to west) of 25 km s$^{-1}$. 
Bright emission around the nucleus has been emphasized in
the `blow-up' of this region shown in Fig.~\ref{fig:features}a.  
The kink is seen at 640 km s$^{-1}$ just to the east (left in the figure)
of the nucleus and the vertical drop continues to the west of the nucleus
where the emission is much fainter since
it falls within the emission gap denoted `C' in
Fig.~\ref{fig:mom0_optical}.  Similar behaviour can be seen in
 Offset $+11^{\prime\prime}$ frame of Fig.~\ref{fig:pv} and this steep gradient
is also hinted at in the CO(J=1-0) data of \cite{gol94b}.
The nuclear gradient is  
delineated by the green line in the figure which has been adopted to
pass through the small emission peak at the nucleus (this peak was pointed out in 
Sect.~\ref{sec:nucleus}).  Our
 modelled ring center (from Table~\ref{tab:galaxy_params}) and our value
of $V_{sys}$ (from the global profile of Fig.~\ref{fig:global_profile})
are marked by the cross and agree, within uncertainties, with the center adopted
for the nuclear gradient.  This
location appears to mark the galaxy's true nucleus and is an additional
argument (see also Sect.~\ref{sec:nucleus}) for adopting the IR center as the
location of the nucleus rather than the minimum at C.


The slope of the nuclear gradient is 4.1 km s$^{-1}$ arcsec$^{-1}$ (94 km
s$^{-1}$ kpc$^{-1}$),
much steeper than the rotation
curve of the central molecular disk,   
implying the presence of 
  a centrally concentrated
mass of M$_{dyn}\,=\,5\,\times\,10^7$ M$_\odot$ within a radius of 
282 pc.
  To our knowledge, this is the first
evidence for a concentration of mass at the center of NGC~4631.


\subsubsection{The Outer Disk and Total Dynamical Mass}
\label{sec:results:outerdisk}

The weak outer disk emission becomes most evident at radii greater than 75$^{\prime\prime}$
(3.3 kpc) and appears kinematically distinct from the central molecular ring
 (see  Fig.~\ref{fig:pv}).  Peak velocities at the
largest measurable radii on either side of
the nucleus are approximately the same as the peak values found in the central molecular
ring, but in the region between the ends of the central molecular ring and
the furthermost radii, velocities are lower, giving the impression that the rotation curve declines
at the ends of the central molecular ring and then rises again with radius.
However, a comparison with PV plots from the HI distribution shows that
the rotation curve does not decline in this region
(see \cite{ran94}, his Fig.~9); rather, it simply becomes
 flat at the ends of the central
molecular ring and outwards.  
Therefore the appearance of lower rotational velocities just outside of the central molecular
ring is a result of
irregularities in the CO(J=3-2) emission
intensity (see next section).  Given the faintness of the CO(J=3-2) emission in the outer disk,
it has not been possible to identify features corresponding to the HI supershells in these
regions (Fig.~\ref{fig:mom0_optical}).

 Taking the mean of the maximum velocities
 ($V_r\,=\,155$ km s$^{-1}$) and radii ($R\,=\,10.6$ kpc) 
 on either side of the
nucleus and adopting a spherical distribution of total (light plus dark) mass for NGC~4631,
the total mass is 
M$_{tot}(R\,<\,11~{\rm kpc})\,=\,6\,\times\,10^{10}$ M$_\odot$.
  The HI distribution reaches comparable velocities
(150 km s$^{-1}$) but can be detected to much larger radius,
i.e. to $R\,\approx\,24$ kpc \citep{ran94}. From HI data, we can therefore
extend the
total mass estimate
to $M_{tot}(R\,<\,24~{\rm kpc})\,=\,1.3\,\times\,10^{11}$
M$_\odot$.  

\subsection{Anomalous Velocity Features and High Latitude Molecular Gas}
\label{subsec:results:faint}

\subsubsection{Anomalous Velocity Features}
\label{subsec:results:supershells}

Fig.~\ref{fig:pv} reveals several anomalous velocity features which appear as extensions
or partial loops in PV space associated with the central molecular ring.   
We have identified five
such features
(labelled F1 through F5 
on the major axis slice) labelling only those that are connected to
the central molecular ring and can be traced over at least two beam sizes spatially, at least
two contiguous velocity channels, and over 2$\sigma$ in intensity.  
Two features, F1 and F3, occur near the ends of the central molecular ring and contribute to
the (inaccurate) appearance of a declining rotation curve in these regions.
To aid in visualizing these features, we have 
spatially smoothed the major axis slice of Fig.~\ref{fig:pv} and show the result in Fig.~\ref{fig:features}b
with the same labelling.  The inset shows the CO(J=1-0) data of
\cite{ran00a} treated similarly.  Although the CO(J=1-0) data show significant noise,
they do reveal some extensions at approximately the same
locations of the features we have identified.  
We have also verified that these features exist in the CO(J=3-2) data cube
that has been independently reduced by \cite{war10} (see Sect.~\ref{sec:obs_datared}).

Some of the labelled features can be traced above or below the major axis.  For example,
the feature, F3, which appears loop-like on the major axis slice of
Fig.~\ref{fig:pv}, shows a split velocity profile
above 
the major axis (Offset = +11$^{\prime\prime}$) open to the east.  This feature
corresponds to the expanding shell observed by \cite{ran00a} in the CO(J=1-0) distribution.
The feature, F2, shows complex structure (Fig.~\ref{fig:features}) but it can be traced south of
the major axis (Offset = -11$^{\prime\prime}$ in Fig.~\ref{fig:pv}) where it has the appearance of
a smaller complete loop.  The feature, F1, could be related to F2, though this association is not
clear.
Feature F4 appears to be an anomalous velocity extension to the main emission,
and F5 forms a large, but weaker loop.

Since these features blend with the main emission in RA-DEC space,
we cannot confirm shell-like structure spatially; however, the appearance of at least  F2 
and F3 in PV space are consistent with known behaviour of expanding shells or
portions of expanding shells such as have been seen 
in our own and other galaxies in HI or CO \citep[e.g.][]{spe04, mcc02,irw96,lee02,wei99}.
 F3 is the first detection 
in CO(J=3-2) of the previously known CO(J=1-0) shell
and the other features are new detections. 


In Table~\ref{tab:anomalous} we provide parameters of these features.
The molecular masses may be lower limits because
CO(J=3-2) emission represents only gas that shows some excitation
and, in addition, we have determined the
 masses only over regions
where the features clearly depart from the main disk emission in
PV space.  Nevertheless, we do find that our mass for F3 agrees with the
value of $\approx\,10^8$ M$_\odot$ measured by \cite{ran00a} for the corresponding
CO(J=1-0) expanding shell.

The velocity width of these
features, $\Delta\,V$, indicates that, even if no shell-like
signature is observed, some expansion $V_{exp}\,=\,\Delta\,V/2$
 must be occurring.  From this information, we estimate 
the kinetic energy of the anomalous velocity features, $E_K$ (see Table~\ref{tab:anomalous}).
These energies may also be lower limits since they scale with mass.
Although we are not certain of the origin of the features (see Sect.~\ref{sec:discussion}),
it is common practice to estimate the input
mechanical energies required to form them,
$E_0$, assuming an instantaneous
energy input, such as would be the case for clustered supernovae and stellar winds over a time
scale that is short in comparison to the age of the feature itself.
The required
mechanical input energy is \citep{che74},
\begin{equation}
\frac{E_{0}}{\rm ergs}\,=\,
5.3\,\times\,10^{43}\,\left(\frac{n_0}{{\rm cm}^{-3}}\right)^{1.12}\,
\left(\frac{{R_{exp}}}{\rm pc}\right)^{3.12}\left(\frac{V_{exp}}{{\rm km~ s}^{-1}}\right)^{1.4}
\label{equation:energy}
\end{equation}
where $n_0$ is the ambient density at the time of the energy 
deposition and $R_{exp}\,=\,D/2$ is the shell radius.
If a continuous energy input is assumed instead,
the results tend to be consistent with Eqn.~\ref{equation:energy} to within a factor
of a few \citep[see][]{spe04}. 
Models such as these typically assume a uniform ambient density 
in the plane which is certainly not
the case for molecular clouds.  For NGC~4631, most
of the molecular mass is in clouds with densities $\approx$ 300 cm$^{-3}$ 
(see Sect.~\ref{sec:results:excitation}).
  Since we estimate that there are roughly equal HI and H$_2$ masses in
the region of the central molecular ring (Sect.~\ref{sec:results:mass}) and
the model of \cite{ran94} estimates the ambient HI density in the plane to be
$0.4$ cm$^{-3}$, then the average H$_2$ density could be similar, leading to
 a volume filling factor of $\approx\,10^{-3}$
for the molecular clouds.   Although these are very rough estimates, most of
the molecular outflow would be through in-plane densities that are lower than those of
the molecular clouds themselves.  We have therefore used 
 $n_0\,=\,0.4$ cm$^{-3}$ in
 Eqn.~\ref{equation:energy}, bearing in mind that the results for $E_0$ will 
be conservative estimates since higher densities could be important at some level. 

Finally, we compute `characteristic' ages,
$\tau$, which do not take into account accelerations or decelerations over the
development of the feature.  If a continuous wind model is adopted, these lifetimes
would decrease by approximately a factor of three \citep{mcc02}.  It is interesting
that all lifetimes fall into a narrow range of timescales,
 $\tau\,\approx\,2\rightarrow 2.6\,\times\,10^7$ yr, 
suggesting that they
may be related to a single burst of star formation.  Note, however, that our
spatial
resolution selects features that are of kpc scale and these observations
would not have detected smaller,
and therefore younger, features, if they were present.  Larger features may also
be difficult to detect if their densities diminish with increasing size.

The CO(J=3-2) shell results of Table~\ref{tab:anomalous} are similar to those of HI shells found in our own
Milky Way 
and external galaxies \citep[e.g.][and others]{bri86,puc92,mcc02, cha01, spe04}.  
The tabulated masses and energies, although order of magnitude estimates,
are likely conservative as discussed above; it
is clear that many hot young stars and supernovae would be required to form the features
if they are indeed the origin.  We will return to this issue in Sect.~\ref{sec:discussion}.

\begin{table*}
 \centering
 \begin{minipage}{200mm}
  \caption{Parameters of the Anomalous Velocity Features}
  \label{tab:anomalous}
  \begin{tabular}{lcccccccc}
\hline
Feature & RA$_0$ $^a$, DEC$_0$ $^a$  &$V_0$ \footnote{Central position and velocity
of the feature, from the centroid of the emission after smoothing spatially and
in velocity\\ using both the major axis slices as well as offset slices (see
Fig.~\ref{fig:pv}) as needed.
 Positional uncertainties are approximately $\pm$ 10$^{\prime\prime}$\\ and the velocity
uncertainty is  $\pm$ 10 km s$^{-1}$. \label{footnote:a}}
 & $D$ \footnote{Diameter of the feature (angular and linear) measured from
the original resolution data.  Uncertainties are as in
Footnote~\ref{footnote:a}.}
& $\Delta\,V$ \footnote{Full velocity extent of the feature measured from the original
resolution data. Uncertainties are as in
Footnote~\ref{footnote:a}. }
 & M$_{mol}$ \footnote{Total molecular mass (including heavy elements),
adopting $R_{3-2/1-0}\,=\,0.47$ (Sect.~\ref{sec:results:mass}); the result is an average
between the\\ smoothed and unsmoothed data.  The uncertainty is $\approx$ $\pm 0.5\,\times\,10^7$ M$_\odot$
which represents a typical flux in the background
over a\\ similar-sized region.} 
& $E_K$ \footnote{Kinetic energy of the feature from $E_K\,=\,(1/2)\,M\,(\Delta\,V/2)^2$.} 
& $E_0$ \footnote{Input energies, from Eqn.~\ref{equation:energy}.}
& $\tau$ \footnote{Characteristic age of the feature, from $\tau\,=\,D/(\Delta V)$.} \\
        & (h m s, $^\circ$ $^\prime$ $^{\prime\prime}$) & (km s$^{-1}$)     
&   ($^{\prime\prime}$, kpc)       & (km s$^{-1}$)      & (10$^7$ M$_\odot$)  & ($10^{53}$ ergs)  
&  ($10^{53}$ ergs)    & ($10^7$ yr)\\
\hline
F1 & 12 42 00, 32 32 15 & 500 & 22, 0.95 & 52 & 2.3  & 1.6 & 4.1 & 1.8\\
F2 & 12 42 03, 32 32 13 & 593   & 44, 1.9 & 73 & 5.3  & 7.1 & 57 & 2.5\\
F3 & 12 42 14, 32 32 40 & 718 & 33, 1.4 & 62 & 9.0  & 8.6 & 18 & 2.2\\
F4 & 12 42 10, 32 32 29 & 593 & 22, 0.95 & 42 & 3.0  & 1.3 & 3.0 & 2.2\\
F5 & 12 42 07, 32 32 25 & 468 & 44, 1.9 & 73 & 2.4  & 3.2 & 57 & 2.5\\
\hline
\end{tabular}
\end{minipage}
\end{table*}

\subsubsection{High Latitude Molecular Gas}
\label{subsec:results:thickness}

As indicated in the previous section, the
 anomalous velocity features seen in the PV plots (Fig.~\ref{fig:pv}) are not
easily traced to high latitudes.  However, we do see some evidence for the presence of
high latitude CO(J=3-2) in the data.

Fig.~\ref{fig:mom0_optical} shows that the
disk thickness of NGC~4631 varies with position, but at least
within the central molecular ring, the evidence suggests that
NGC~4631 forms a thick, rather than a thin distribution.  
A thin global molecular gas disk with the observed diameter of the
 CO(J=3-2) emission (see Sect.~\ref{sec:results:outerdisk}) 
which is inclined by 86$^{\circ}$
 (Table~\ref{tab:galaxy_params}) 
would project to a total 
apparent vertical extent of only 48$^{\prime\prime}$ including the smoothing effects of
the beam,
whereas the total observed minor axis extent
 is approximately 65$^{\prime\prime}$ (2.3 kpc after beam correction, or $z\,=\,1.2$ kpc).
 If the inclination of the central molecular ring
were as low as 83$^\circ$, then it could be interpreted as thin.  
However, our model of the central molecular ring (Table~\ref{tab:ring_model}) gives best results
for an even higher inclination ($i\,=\,89^\circ$); lower values are poorer. 

\begin{figure}
\includegraphics[scale=0.35]{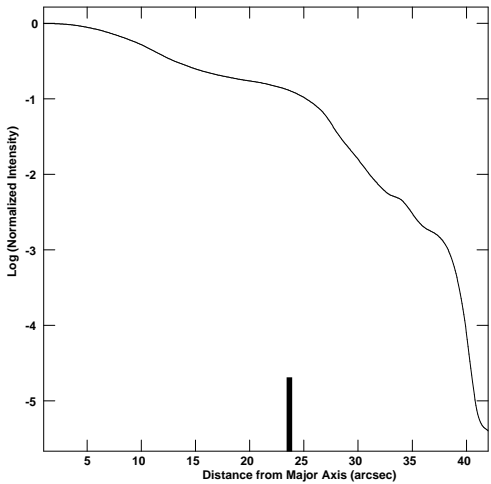}
\caption{Vertical profile of the CO(J=3-2) emission of NGC~4631, formed from
Fig.~\ref{fig:mom0_optical} averaged over a 100$^{\prime\prime}$ wide region centered at
the nucleus.  The profiles to the north and south have also been averaged to form a
sensitive plot.  The emission cuts off at $z\,\approx\,40^{\prime\prime}$ due to the
algorithm for determining the total intensity image (see caption of
Fig.~\ref{fig:mom0_optical}).  The projected minor axis extent of a thin disk of radius, 
$R\,=\,243^{\prime\prime}$, inclined at $i\,=\,86^\circ$ ($R\, cos(i)$) and then
smoothed by the beam, is given by the thick vertical line segment.}
\label{fig:minoraxis}
\end{figure}

To further investigate the vertical extent of the molecular gas, we have formed a 
plot at high sensitivity showing the minor axis profile by averaging over a 
100$^{\prime\prime}$-wide region of the major axis
(i.e. approximately over the FWHM of the central molecular ring) and then averaging the
north/south sides.  The result, 
shown in Fig.~\ref{fig:minoraxis}, reveals a complex profile which is not well described
by a single smooth fit.  CO(J=3-2) can be traced out to $z\,=\,33^{\prime\prime}$ (1.4 kpc),
correcting for beam smoothing.  Thus, the central molecular ring of NGC~4631
appears to have a thick vertical distribution of CO. 
   Given the known
halo activity in this galaxy,
most of which has been measured to much larger values of $z$ 
(see Sect.~\ref{sec:intro}),
 the presence of thick, agitated molecular gas is perhaps not
surprising.  
For comparison, we note that main sequence stars in NGC~4631 have been measured out to a $z$ height
of 2.3 kpc and AGB and RGB stars have been measured to even higher $z$ values 
\citep{set05b}.  

We see no evidence from the PV plots (Fig.~\ref{fig:pv}), however,
for
a change in the slope with $z$ height (`lagging halos') such as has been seen in
HI or ionized extraplanar gas in several other galaxies
\citep[e.g.][]{tul00,ran00b,fra02,oos07} although the vertical extent of
 molecular gas in NGC~4631 is, in general, smaller than these other components.

Fig.~\ref{fig:mom0_optical} also shows a number of disconnected emission features
 above
and below the plane at distances of $50^{\prime\prime}\,-\,60^{\prime\prime}$
($z \,\approx\,2.2\,\rightarrow\,2.7$ kpc). 
The smoothing and noise cut-off techniques used to create the total intensity image
are ideal for emphasizing such low level emission which generally can only be seen in
cubes that are smoothed, rather than in the original channel maps.  Their emission 
is contiguous in velocity space and
several can be identified in the independently reduced, lower velocity resolution 
data of \cite{war10}.  However, since not every one can be independently confirmed,
we do not label each individually and only consider the energetics (below) of a `typical'
feature. 
 For the purpose of discussion,
we refer to them as high latitude `clouds'.

\begin{figure}
\includegraphics[scale=0.49]{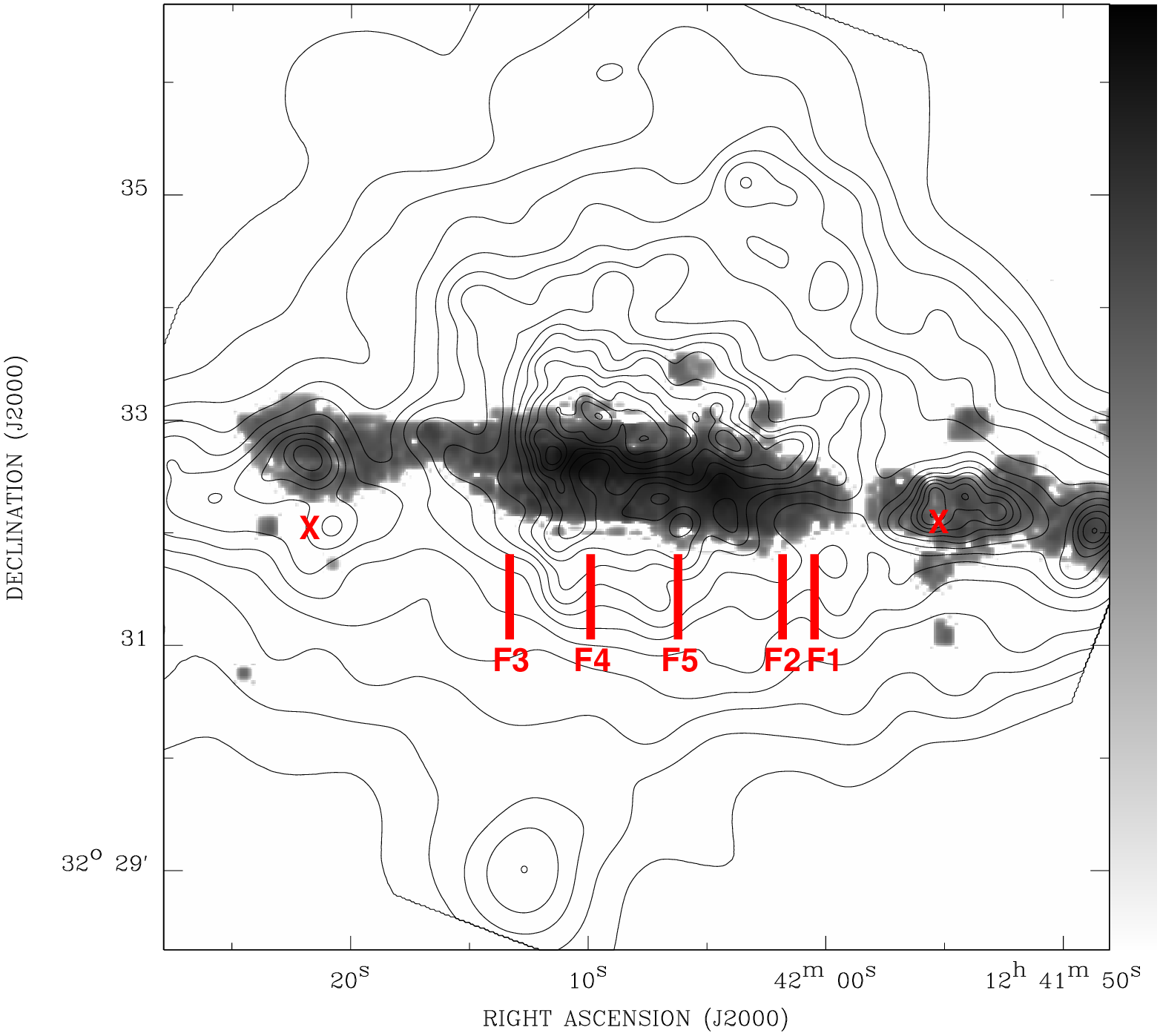}
\caption{CO(J=3-2) map of Fig.~\ref{fig:mom0_optical} with low level emission emphasized
overlaid with X-ray contours representing hot gas (from the data of
Wang et al. 2001).  Approximate locations of the features identified in
Fig.~\ref{fig:pv} are marked as well as the two HI shells (marked with X) identified
by Rand \& van der Hulst (1993).}
\label{fig:wangs_overlay}
\end{figure}

Fig.~\ref{fig:wangs_overlay}, which emphasizes the low level emission of
Fig.~\ref{fig:mom0_optical}, shows how these clouds exist above apparent disturbances in the disk.
For example, there are clouds both  
above and below the location of
the western HI supershell (the western X).  There is also a cloud above the plane and
a 
 complete loop
below the plane at the location of feature F5.
These clouds
 have masses of 
M$_{cl}\approx\,10^6$
M$_\odot$ and estimated
potential energies of,
\begin{eqnarray}
\frac{\Phi(z)}{\rm ergs}\,=\,2\,\times\,10^{54}\,\left(\frac{{\rm M}_{cl}}{2\,\times\,10^7~{\rm M}_\odot}\right)\,
\left(\frac{z_0}{700 ~{\rm pc}}\right)^2\,\nonumber\\
\cdot\left(\frac{\rho_0}{0.185\,{\rm M}_\odot~{\rm pc}^{-3}}\right)
\times\,ln\left[cosh\left( \frac{z}{z_0} \right)\right]
\end{eqnarray}
\citep[e.g.][]{spe04}, where $z_0$ is the mass scale height and $\rho_0$ is the mid-plane mass
density with the normalization factor taken to be an estimate for the local Solar neighbourhood.
Adopting the stellar scale height of $z_0\,=\,1$ kpc from \cite{set05b} we find,
\begin{equation}
\frac{\Phi}{\rm ergs}\,\approx\,3.7\,\times\,10^{53}\,
\left(\frac{\rho_0}{0.185\,{\rm M}_\odot~{\rm pc}^{-3}}\right)
\end{equation}
which suggests that kinetic energies of this order are required to place the molecular
clouds at their observed heights.  Input energies are likely higher
still (cf. $E_K$, $E_0$ in Table~\ref{tab:anomalous}) if these clouds have originated in
the disk.  Further discussion can be found in Sect.~\ref{sec:discussion}.

\subsubsection{Comparison of Outflow and High Latitude Features with other Wavebands}

As indicated in Sect.~\ref{sec:intro}, the halo of NGC~4631 has been observed in
every ISM component.  Since
the halo is so extensive and there is overlap with the broader HI tidal streamers,
some caution must also be exercised in identifying correlations as they may
be due to 
 chance projections along the line of sight \citep[see e.g.][]{tay03}.
The features
that we have identified (Table~\ref{tab:anomalous}) are mostly visible in PV space, rather
than in RA-DEC and we have thus not found counterparts at other wavebands, other than
F3 as noted in Sect.~\ref{subsec:results:supershells} which is associated with a
 CO(J=1-0) expanding shell.


A possible high latitude dust arch, discovered by \cite{nei99}, but interpreted as
part of HI tidal spur 4 by \cite{tay03}, has its footprint in the disk at approximately
the position of the west HI expanding shell. At this location, early X-ray images show
hot gas extending into the halo, as shown most clearly in \cite{wan95} (their Fig.~1a).
More recent X-ray images (see Fig.~\ref{fig:wangs_overlay}) show 
emission from hot gas which has originated from the disk 
and is associated with H$\alpha$ emission and a `froth' of superbubbles
\citep{wan01}.  Given the varying distributions of the X-ray and CO(J=3-2) components and
the edge-on nature of the galaxy,   
Fig.~\ref{fig:wangs_overlay} does not show clear correlations between the two components.
However, it is clear that the stronger X-ray halo emission is located over the region of
the central molecular ring.

Finally, one high latitude CO(J=3-2) feature, namely the southern loop associated with F5
(Fig.~\ref{fig:wangs_overlay}),
does appear to have an H$\alpha$ counterpart in the form of two H$\alpha$ spurs, 
the latter seen in Fig.~\ref{fig:overlays}a. To the east of the two spurs is a third H$\alpha$
spur that appears to have a counterpart in the $\lambda\,160$ $\mu$m emission 
(Fig.~\ref{fig:overlays}c).


\section{Discussion}
\label{sec:discussion}

A picture has emerged of NGC~4631 as an actively star forming galaxy, but not one with
a central starburst.  
Approximately 80\% of the massive star formation in this galaxy occurs outside
of the central 1.8$^\prime$ (4.7 kpc) diameter region
and H$\alpha$ and UV emission in the disk are also widely distributed.   
The X-ray halo (Fig.~\ref{fig:wangs_overlay}), which resembles the well-known radio halo
\citep{wan01},
 is similarly widespread over the entire star forming disk.   Widespread halo emission
is seen in all ISM components (see Sect.~\ref{sec:intro} for references) and
all evidence, including the presence of HI supershells, a CO(J=1-0) expanding shell,
vertical filaments, the vertical
structure of the magnetic field lines, the concentration of stronger halo emission over
stronger star forming regions in the disk, and now CO(J=3-2) anomalous velocity features
point to 
star forming regions and their related supernovae
and stellar winds in the disk as the origin of the halo. The question is,
why does NGC~4631, whose halo is arguably the most spectacular ever observed,
 have such a wide-spread, prominent halo in comparison to other edge-on galaxies?

To consider this question, it is worth comparing NGC~4631 to two other galaxies:
NGC~5775, which
has a similarly wide-spread multi-phase halo \citep{lee01,li08}, and
M~82, which is the prototypical nuclear starburst with bipolar outflow.
All three galaxies have significant companions with which they are interacting, the 
most obvious evidence being the presence of 
HI bridges or streamers between them \citep{yun94,
irw94, wel78}.  For NGC~4631, the interaction is with the large spiral, NGC~4656 and the
dwarf elliptical, NGC~4627 (Sect.~\ref{sec:intro}).  
These interactions may have played an important role in forming their halos
via tidal disruptions or
 agitation of the disk.  For example, if the disk has become `puffed up', such as we seem
to see in the thick CO(J=3-2) layer in NGC~4631 (Sect.~\ref{subsec:results:thickness}),
 it would be easier for 
 material to escape into the halo.  
Stellar winds and supernovae should also produce thickened gaseous disks as material escapes vertically
into the halo.
For NGC~4631, since the stellar disk also extends several kpc above the plane
(Sects.~\ref{sec:intro},~\ref{subsec:results:thickness}), the interaction 
has most likely played an important direct role.


As for star formation, M~82, NGC~5775 and NGC~4631
all have similar FIR luminosities to within a factor of about 2,
but their luminosities per unit optical disk area, a distance-independent
quantity, vary in the ratio, 1:0.16:0.05 in the order listed above
\citep{tul06a}.
NGC~5775 and NGC~4631, which both show disk-wide halos, differ by only a factor of 3.
M~82, on the other hand, has a SFR per unit area
that is higher than that of NGC~4631 by a factor of 20. 
Since M~82 has been well-studied, it is possible to compare activity within
the nuclear region itself.  For M~82,  the 
supernova rate in the central 700 pc  is  $\nu\,=\,0.1$ yr$^{-1}$
\citep{kro81}. From Eqn.~\ref{equation1}, 
Fig.~\ref{fig:ratio_mass}d, and relations in \cite{con92}, we find 
$\nu\,=\,0.001$ yr$^{-1}$ for an equivalently sized region at the center of NGC~4631.  
That is, the central supernova rate in NGC~4631 is two orders of magnitude smaller than in M~82, 
hence the nuclear outflow in M~82 but not in NGC~4631.

It is well known that interactions can  
trigger a build-up of molecular gas in galaxy centers and can also induce strong central starbursts
 \citep[e.g.][]{aal07}. 
 What is particularly striking about NGC~4631 is
the presence of strong central molecular emission 
in a ring out to a radius of $\approx$ 5 kpc (e.g. Fig.~\ref{fig:major_axis}) but 
without a central starburst.
The physical properties of the ring are typical of galaxy disks, rather than other known
nuclear starbursts (Sect.~\ref{sec:results:excitation}) and the star formation efficiency is 
not very high;
that is, if SF continues at the
current rate, there is sufficient molecular gas to sustain it for at least $3\,\times\,10^9$ yr.
Inside the ring, within the 
central 17$^{\prime\prime}$ (740 pc) diameter region (Fig.~\ref{fig:ratio_mass}) the SFR is 
enhanced and there is a peak in the hot dust distribution
(Fig.~\ref{fig:slices}) as well as a small peak in CO(J=3-2) (Fig.~\ref{fig:major_axis}).
Even here, however, the gas consumption time scale is
at least $10^9$ yr.  If the interaction is to trigger a central starburst in NGC~4631, then it
has already happened or it has not happened yet.

\cite{kna09} have added further confirmation to the notion that, 
although bursts of SF can occur as a result
of interactions and there is a tendency for the star formation to be centrally 
concentrated,  interactions can also initiate continuous star formation over
longer timescales, ie. a few $\times\,10^{8}$ yrs 
\citep[see also][]{dim08}.  For NGC~4631, the 
interaction with its companion occurred
$\approx\,3\,\times\,10^8$ years ago \citep{com78}.
There is some evidence that at least one
burst of star formation (or higher SFR) 
has already occurred in NGC~4631.
From UV observations \cite{smi01} note
that the {\it current} rate of SF does not seem to account for other indications of strong
outflow in this galaxy.  For example, the eastern HI shell (see Fig.~\ref{fig:mom0_optical}) can be
explained by supernovae from a massive SF region containing $5.3 \,\times\,10^4$ OB stars beginning about 
$2\,\times\,10^7$ yr ago and
that the currently observed UV emission (which is insufficient to explain the shell)
 is due to second generation stars.  
Our own estimates for the lifetimes of the observed CO(J=3-2) anomalous velocity features are 
approximately the
same (Table~\ref{tab:anomalous}) and, estimated the same way, the HI supershells observed by
\cite{ran93} also result in ages of a few $10^7$ yr. These results suggest that a higher
rate of SF may have occurred of order $\approx\,10^7$ yr ago.  However, as noted in 
Sect.~\ref{subsec:results:supershells}, selection effects may have prevented us from detecting
features with larger or smaller lifetimes, so we cannot place a limit
on the timescale for past SF in general.


Estimates for 
the kinetic energies of the anomalous velocity features are of order $E_K\,\approx\,10^{53}$ ergs and estimates of
potential energy of the small clouds above and below the plane are about the same
(Sect.~\ref{subsec:results:thickness}).  The implied input energies are higher still,
possibly by an order of magnitude (Table~\ref{tab:anomalous}), a conclusion also implied  
for the energetics of the outflowing molecular gas in the nuclear outflow of M~82 
\citep{sea01}.  These energies
 are also typical of what has been previously seen for HI and CO
expanding shells in other galaxies (Sect.~\ref{subsec:results:supershells}) and are quite high.
 If the observed features are associated with a higher SFR in the past
as suggested above, however, the energies may be sufficient.  Supernovae from 
the stars that are believed to be responsible
for the eastern HI supershell could supply (at 10$^{51}$ ergs each) 
 $5.3\,\times\,10^{55}$ ergs for this shell which is more than adequate.  
 However, there would need to be at least five such SF regions throughout the region of the central
molecular ring to account for all of the features of Table~\ref{tab:anomalous}.  There is still some need for 
time-dependent modelling of such outflows
in a realistic multi-phase, multi-density, and magnetised ISM.

The prominent radio halo 
 in NGC~4631 actually reflects an integration over past SF
activity in the disk.  The average magnetic field strength in the disk of NGC~4631 is 
$B\,=\,6.5$ $\mu$G \citep{dah95} with a CR electron lifetime estimate of
$t_{CR}\,=\,4.8\,\times\,10^7$ yrs.  The ratio between the halo and disk magnetic field
strength is 5/8 \citep{hum91} yielding an average halo field strength of
$B\,=\,4$ $\mu$G and, since $t_{CR}\,\propto\,B^{-3/2}$, the CR lifetime in the halo is
$t_{CR}\,\approx\,10^8$ yrs.  Consequently, the radio halo that we see today has a memory
of outflows that have occurred since approximately the time that the interaction occurred,
assuming that the outflows have not escaped into the intergalactic medium.  Whether or
not this assumption is correct
requires deeper and more extensive mapping of the magnetic field direction than is currently
available.  However, arguments presented in \cite{sea01} suggest that even in the 
more energetic nuclear outflow
of M~82, at least the molecular gas and dust do not escape.
 
The implication of a higher SFR in the past suggests that, when strong halos such as in
NGC~4631 and NGC~5775 are observed, they may have been enhanced by a 
previous outflow event (or events).
Although rather
speculative, another possibility is that such galaxies have also experienced a stronger
 M~82-like starburst and nuclear outflow in the past that was triggered by the interaction.
In the case of NGC~4631, the enhanced SFR right at the nucleus could be a remnant of a
past `M~82-like' nuclear starburst.
We could imagine the
interior of the nuclear molecular ring 
to have been excavated by massive star formation and outflow.
The minimum near the nucleus at `C' (Fig.~\ref{fig:mom0_optical}) could reflect local variations in SFR
and molecular content.  The past additional `boost' of outflow material into the halo may be what is required to
distinguish between spectacular halos and those that are more modest.

\section{Conclusions}
\label{sec:conclusion}

As part of the JCMT Nearby Galaxies Legacy Survey, we
have mapped the CO(J=3-2) emission from the edge-on galaxy, NGC~4631, which is known for its spectacular
multi-phase gaseous halo.  

Most of the CO(J=3-2) emission is concentrated 
within a radius of  $\approx$ 5 kpc.  Although the spatial distribution could be more
complex, the emission 
 is well modelled by a simple edge-on
 ring with a Gaussian density distribution which 
peaks at a radius of 1.8 kpc and has inner and outer scale
lengths of 0.1 and 0.28 kpc, respectively.
The center of the ring agrees with the IR center.  A small CO(J=3-2) peak occurs within
this ring, right at the nucleus.
Outside of the central molecular ring, weaker more extensive disk emission 
is present which has been mapped out as far as
9.25 kpc to the east of the nucleus and 12.4 kpc to west.   This radial extent exceeds that of any previous
CO observations. 

Comparisons have been made between CO(J=3-2), $\lambda\,24~\mu$m,
$\lambda\,160~\mu$m emission and H$\,\alpha$.
We find that the CO(J=3-2) emission more
closely follows $\lambda\,24~\mu$m (a hot dust tracer) rather than
$\lambda\,160~\mu$m emission  (a cold dust tracer), suggesting that CO(J=3-2) is a good
tracer of star formation.
 H$\,\alpha$ emission
is uncorrelated because of the high extinction in this edge-on galaxy. 

 For the inner 2.4 kpc radius region of the central molecular ring, we have formed the
first spatially resolved 
maps of $R_{3-2/1-0}$, the H$_2$ mass surface density, and the SFE
for NGC~4631.  Only 20\% of the global SF
occurs in this region.
We find that $R_{3-2/1-0}$ is typical of galaxy disks, in general, rather than of regions associated
with central starbursts.  Molecular cloud densities  ($\approx\,10^3$ cm$^{-3}$)
in this region are also typical of
molecular clouds in galaxy disks rather than central starbursts. 
 The SFE in this region is, on average, $6.4\,\times\,10^{-10}$ yr$^{-1}$ leading to a mean
gas consumption timescale of $2.6\,\times\,10^9$ yr for the H$_2$ and longer if HI is
included.  That is, the SFR in this region is modest when compared to the abundant gas that is present.
There is, however, an enhanced SFR and SFE right at the nucleus (within a central region of
740 pc diameter), although the gas consumption timescale is still long ($10^9$ yr).

The total
molecular gas mass in NGC~4631 is  M$_{H_2}\,=\,2.2\,\times\,10^9$ M$_\odot$, which improves upon
previous values.  Since the total HI mass is M$_{HI}\,=\,1.0\,\times\,10^{10}$ M$_\odot$, the total gaseous
content of the galaxy is dominated by HI.  The global gas-to-dust ratio is 170.

The velocity field of NGC~4631 is dominated by the steeply rising rotation curve of
the central molecular ring followed by the flatter outer disk; the peak rotational velocity is 
$V\,=\,155$ km s$^{-1}$.
The total dynamical mass within 11 kpc radius 
is $6\,\times\,10^{10}$
M$_\odot$.  
At the center of the galaxy, we find a steep rotation curve, providing the
first evidence for 
a central concentration of mass, i.e.
M$_{dyn}\,=\,5\,\times\,10^7$ M$_\odot$ within a radius of 282 pc.


We can now add CO(J=3-2) emission to the long list of evidence for outflowing gas in NGC~4631.  
Five anomalous
velocity features with properties 
similar to those found in expanding shells (or parts thereof)
in other galaxies have been detected, all associated with the central molecular ring.  One of these (F3) 
corresponds to an expanding CO(J=1-0) shell previously found by \cite{ran00a}.  The galaxy also has a thick
CO(3-2) disk which we trace to a $z$ height of 1.4 kpc.  Some small `clouds' 
are observed at higher latitudes,
possibly associated with outflows from the disk.  We suggest a scenario in which interactions
with the companion galaxies in the past has produced enhanced star formation throughout the disk
and 
speculate that there could have been a massive nuclear outflow in the past.

\section*{Acknowledgements}
\label{sec:acknowledgements}
We are grateful to R. Rand for supplying the BIMA CO(1-0) data and to R. Wielebinski
and M. Krause for supplying IRAM CO(1-0) data for comparison purposes.
This research has made use of the NASA/IPAC Extragalactic Database (NED) which is operated by the 
Jet Propulsion Laboratory, California Institute of Technology, under contract with the National 
Aeronautics and Space Administration. The lead author wishes to thank the Natural Sciences and
Engineering Research Council of Canada for a Discovery Grant.


\bsp

\label{lastpage}

\end{document}